\documentclass[11pt]{article}
\usepackage[a4paper,margin=2.2cm]{geometry}
\usepackage[T1]{fontenc}
\usepackage{lmodern,amsmath,amssymb,amsthm,array,longtable,booktabs,graphicx,microtype,xcolor}
\usepackage[numbers,sort&compress]{natbib}
\usepackage[colorlinks=true,allcolors=teal]{hyperref}

\newcommand{\Ncarbon}{N_{\mathrm C}}

\title{From quasiperiodic three-colour order on a triangular lattice to covalent carbon sheets}
\newcommand{\manuscriptauthor}{%
  \parbox{0.93\textwidth}{\centering
    {\large Chaoyu He\textsuperscript{1,2}}\par\smallskip
    {\small \textsuperscript{1}School of Physics and Optoelectronics, Xiangtan University, Xiangtan 411105, China\par}
    {\small \textsuperscript{2}Laboratory of Computational Condensed Matter Physics and Materials Quantum Engineering of Hunan Provincial Universities, Xiangtan University, Xiangtan 411105, China\par}
    \smallskip
    {\small Corresponding author: \href{mailto:hechaoyu@xtu.edu.cn}{hechaoyu@xtu.edu.cn}\par}%
  }%
}
\author{\manuscriptauthor}
\date{}

\newtheorem{theorem}{Theorem}

\begin{document}

\maketitle
\begin{abstract}
Can the quasiperiodic colouring of a lattice be encoded in the bonds of a single carbon sheet? We construct golden-, silver- and bronze-mean three-colour patterns on a triangular lattice. The positions of one colour determine the other two through a nearest-neighbour rule. The three patterns have eight, nine and seven first-shell environments, respectively. A triangle of three identical motifs frustrates the alternating bond ports of each of the three molecular alphabets in the golden and silver patterns. The bronze pattern admits four designs with direct or ethynylene connections. Three periodic structures reproduce its infinite environment frequencies in a local-energy model. For a 708-atom sheet with independently matched environment counts, the predicted energy lies 0.295 meV per carbon below the density-functional result. Across four periodic cells initialized from two smaller donors, relaxation changes the bond lengths by at most 0.00113~\AA\ in root-mean-square terms. Fitting one $60^\circ$ sector to relaxed periodic geometries gives complete sixfold-symmetric fragments with up to 22,140 carbon atoms. In the 13,950-atom directly linked sheet, a $\pi$ doublet places 79.05\% of its weight in a boundary shell comprising 7.96\% of the atoms, whereas the lowest unoccupied doublet places only 0.124\% there. These calculations connect a testable motif-selection rule to the energetics and finite-patch spectra of candidate quasiperiodic carbon sheets.
\end{abstract}

\section{Introduction}
Quasicrystals established that long-range structural order need not be generated by translations of a unit cell\cite{Shechtman1984,Levine1984,Levine1986}. Inflation rules and higher-dimensional constructions make this departure from periodicity precise, linking local patterns to irrational proportions and hierarchies of length scales. The range of such structures extends beyond the familiar fivefold Penrose geometry\cite{Penrose1974}: hexagonal bronze-mean order, aperiodic hexagonal tilings and sequence-modulated triangular networks reveal distinct ways to organize nonperiodic matter\cite{Dotera2017,Socolar2011,Matsubara2024,Matsubara2025}. For electrons, this structural freedom raises a direct question: which properties are controlled by a finite catalogue of local environments, and which require the extended hierarchy?

Carbon is a compelling setting for this question because its bonding supports a broad range of connected networks. The experimental biphenylene network, amorphous carbon monolayers and graphdiyne films illustrate the chemical reach of nonhexagonal rings, disorder and acetylenic connectivity\cite{Fan2021,Toh2020,Li2010}. Carbon quasicrystals also have established precedents: $30^\circ$ twisted bilayer graphene exhibits genuine quasicrystalline order and electronic coupling\cite{Ahn2018,Yao2018,Yu2019}, and diamane quasicrystals have been proposed\cite{Diamane2022}. Molecular self-assembly supplies another route to two-dimensional quasiperiodic organization\cite{Molecular2014}. The distinct design problem addressed here is to encode a prescribed aperiodic order in the \emph{in-plane covalent connectivity of a single carbon network}, rather than to rely on the relative orientation of periodic layers. A proposed geometry must satisfy both global order and local valence constraints; neither condition follows from a visually quasiperiodic pattern alone.

A finite local catalogue suggests a computational strategy, but also imposes a limitation. Small periodic networks can reproduce selected local statistics even when none reproduces the complete infinite structure. Locality and cluster-expansion ideas motivate an energy representation in terms of these statistics\cite{Prodan2005,vandeWalle2002}; they do not make the electronic energy exactly additive. Conversely, agreement of local energies does not imply agreement of spectra. A periodic band dispersion and a finite aperiodic spectrum answer different questions and require separate calculations. This distinction allows an energy approximation to be tested without replacing the central quasiperiodic object by an arbitrary periodic analogue.

We ask which of three metallic-mean colourings can be written into the connectivity of a single carbon monolayer. The construction puts the quasiperiodic information into a set of A sites. B and C record their surroundings. A local orientation constraint then selects the colourings that three carbon-fragment alphabets can decorate. For the compatible bronze hierarchy we test energy transfer with a separately relaxed large cell. We then use the same periodic basis to construct sixfold-symmetric open fragments and determine which of their frontier states concentrate at the boundary.

\section{Three metallic means on the same triangular lattice}
A common recursion generates a set of candidate sites. At the first step take $(p,q)=(1,m)$, with $m=1,2,3$ for gold, silver and bronze. Every subsequent step replaces $(p,q)$ by $(q,mq+p)$ and copies the entire previous point set. Place a copy at the origin, one at distance $q-p$ along each of the six lattice directions, and one at each position reached by taking $q$ steps along a direction and $p$ along either adjacent direction. The ratio $q/p$ approaches $\tau_m=(m+\sqrt{m^2+4})/2$. The first three pairs are $(1,1),(1,2),(2,3)$ for gold, $(1,2),(2,5),(5,12)$ for silver and $(1,3),(3,10),(10,33)$ for bronze.

\begin{figure}[p]\centering
\includegraphics[width=\linewidth]{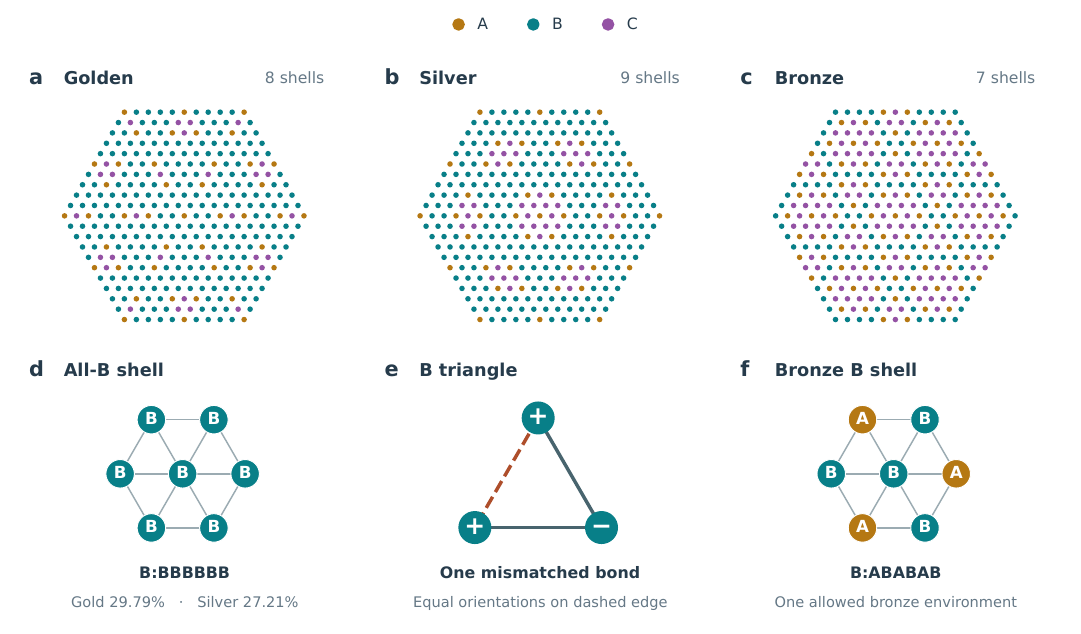}
\caption{\textbf{Three quasiperiodic colourings and a local bonding obstruction.} The upper panels use the same triangular-lattice window and colour key. The B-centred shell in (d) contains an elementary B triangle. In (e), the two signs are the two orientations of the alternating B ports. The dashed bond joins equal orientations in the assignment shown. Reversing the top sign moves the mismatch to the other upper bond. Panel (f) shows one of the two B-centred bronze environments.}\label{fig:family}
\end{figure}

Take the union of the candidate sites over all levels. Retain each isolated candidate as A. For a candidate with a candidate nearest neighbour, require all three track numbers $|i|$, $|j|$ and $|i+j|$ to belong to $K_m=\{\operatorname{round}(k/d_m): k=0,1,\ldots\}$, with $d_m$ the distance from $\tau_m$ to the nearest integer. This last selection is a construction choice. The bronze candidates are all isolated. In gold and silver the candidates cover the lattice and the track test selects the A sites (SI). On every remaining site place C if two opposite nearest neighbours are A, and B otherwise. Thus the same local rule reconstructs the three-colour field from the binary A-site pattern for all three means. Figure~\ref{fig:family} compares equal windows of the resulting fields.

Their first-shell catalogues contain eight, nine and seven environments. The decisive chemical constraint is visible on one elementary triangle. The gold and silver patterns contain a B site with six B neighbours at respective frequencies of 29.79\% and 27.21\% of all lattice sites. Take the centre and any two consecutive neighbours. Each B fragment in the three proposed alphabets has alternating single and double ports and two distinguishable orientations. A matched B--B bond requires opposite orientations. Assigning opposite orientations along the first two edges leaves equal orientations at the ends of the third edge. This excludes the three tested alphabets for gold and silver under the stated matching rule. The obstruction concerns these carbon motifs; other motifs define a different assembly problem.

\section{The chemically compatible bronze hierarchy}
Let $\mathbf e_1=(1,0)$ and $\mathbf e_2=(1/2,\sqrt3/2)$ define the triangular lattice. Integer coordinates have six neighbour directions
\begin{equation}
U=\{(1,0),(0,1),(-1,1),(-1,0),(0,-1),(1,-1)\}.
\end{equation}
Define
\begin{equation}\label{eq:pq}
p_0=1,\quad q_0=3,\qquad p_{h+1}=q_h,\quad q_{h+1}=3q_h+p_h.
\end{equation}
The scale ratio approaches the bronze mean $\beta=(3+\sqrt{13})/2$. With $\alpha=(3-\sqrt{13})/2=-\beta^{-1}$,
\begin{equation}
p_h=\frac{\beta^{h+1}-\alpha^{h+1}}{\sqrt{13}},\qquad
q_h=\frac{\beta^{h+2}-\alpha^{h+2}}{\sqrt{13}}.
\end{equation}
At level $h$, take the 19 offsets
\begin{equation}\label{eq:digits}
D_h=\{0\}\cup\{(q_h-p_h)u_k\}_{k=0}^{5}
\cup\{q_hu_k+p_hu_{k\pm1}\}_{k=0}^{5},
\end{equation}
where the indices are cyclic. Starting with $S_0=\{0\}$, form the set sums $S_{h+1}=S_h+D_h$, removing duplicates, and let $S_\infty=\bigcup_h S_h$. A site is coloured A if it belongs to $S_\infty$; a remaining site is C if it lies midway between two opposite nearest-neighbour A sites, and B otherwise. This is a three-letter labelling, not a proper graph colouring: BB and CC edges are allowed.

\begin{figure}[p]\centering
\includegraphics[width=\linewidth]{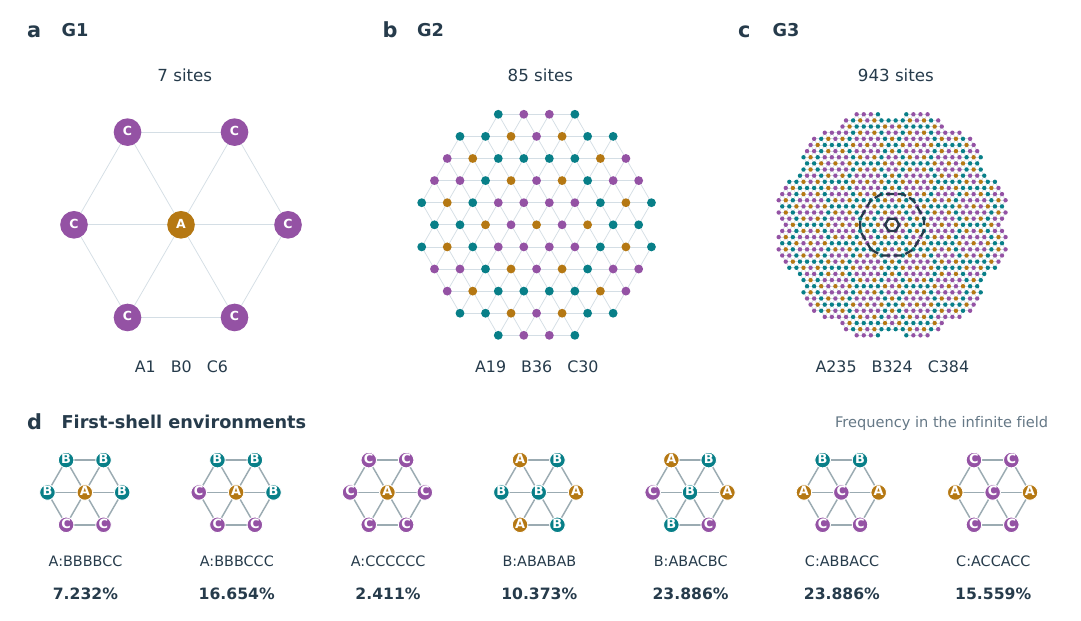}
\caption{\textbf{A triangular three-letter hierarchy.} G1, G2 and G3 contain 7, 85 and 943 molecular sites. Solid and dashed outlines on G3 mark the convex envelopes of the chosen G1 and G2 supports; they show the generation relationship, not artificial grain boundaries. The lower row gives all seven first-shell environments and their infinite frequencies, rounded to the displayed precision. The finite G3 composition is not substituted for the infinite limit.}\label{fig:hierarchy}
\end{figure}

The prescription is locally computable despite its infinite union. With $\rho(i,j)=\max(|i|,|j|,|i+j|)$ and $R_h=\sum_{k<h}(p_k+q_k)<2p_h$, a sum whose highest nonzero digit is at level $h$ has norm at least
\begin{equation}
H_h=q_h-p_h-R_h,\qquad H_{h+1}=H_h+p_h\longrightarrow\infty.
\end{equation}
Thus only finitely many recursion levels affect any fixed observation window. The carbon samples G1, G2 and G3 are prescribed complete-ABC supports of this common field, containing respectively $(1,0,6)$, $(19,36,30)$ and $(235,324,384)$ A/B/C sites (Fig.~\ref{fig:hierarchy}). Their labels specify finite supports; they are not interchangeable with the digit index $h$.

The order can be established without relying on appearance. On a lattice axis the recursion is coded by the primitive interval substitution
\begin{equation}\label{eq:sub}
X\mapsto XYX,\qquad Y\mapsto XZX,\qquad Z\mapsto XZXZX,
\end{equation}
with base words $X=\mathrm{AC}$, $Y=\mathrm{ABB}$ and $Z=\mathrm{ABBBB}$. Its A density is $2\beta/(5\beta+1)$, which is irrational; the axis word cannot be periodic. Sixfold invariance then excludes every nonzero planar translation. In addition, a closed 301-state collared-rectangle representation is primitive and carries growing collars; it implies that each finite coloured patch recurs with bounded gaps. The SI gives the construction, integer certificate and recurrence proof. These statements establish aperiodicity and uniform recurrence in real space. A pure-point diffraction theorem for the relaxed atomic networks is a separate assertion and is not inferred from these results.

The same certificate closes the first-shell catalogue at exactly seven environments. Each environment includes the central letter and the cyclic word of its six neighbours, with rotations and reflections identified. In the order displayed in Fig.~\ref{fig:hierarchy}, their limiting frequency vector is
\begin{equation}\label{eq:f}
\mathbf f_\infty=\frac{(6,\,6(\beta-1),\,2,\,2(\beta+1),\,6\beta,\,6\beta,\,3(\beta+1))}{23\beta+7}.
\end{equation}
Summing by centre gives
\begin{equation}\label{eq:abc}
(f_A,f_B,f_C)=\frac{(2,\sqrt{13}-1,3)}{4+\sqrt{13}}
\simeq(0.262966,0.342585,0.394449).
\end{equation}
The seven first-shell environments should not be confused with pair interfaces: only AB, AC, BB, BC and CC occur. Chemical port orientations can resolve additional distinctions beyond either catalogue.

\section{From a molecular code to four carbon networks}
\begin{figure}[p]\centering
\includegraphics[width=.94\linewidth,height=.81\textheight,keepaspectratio]{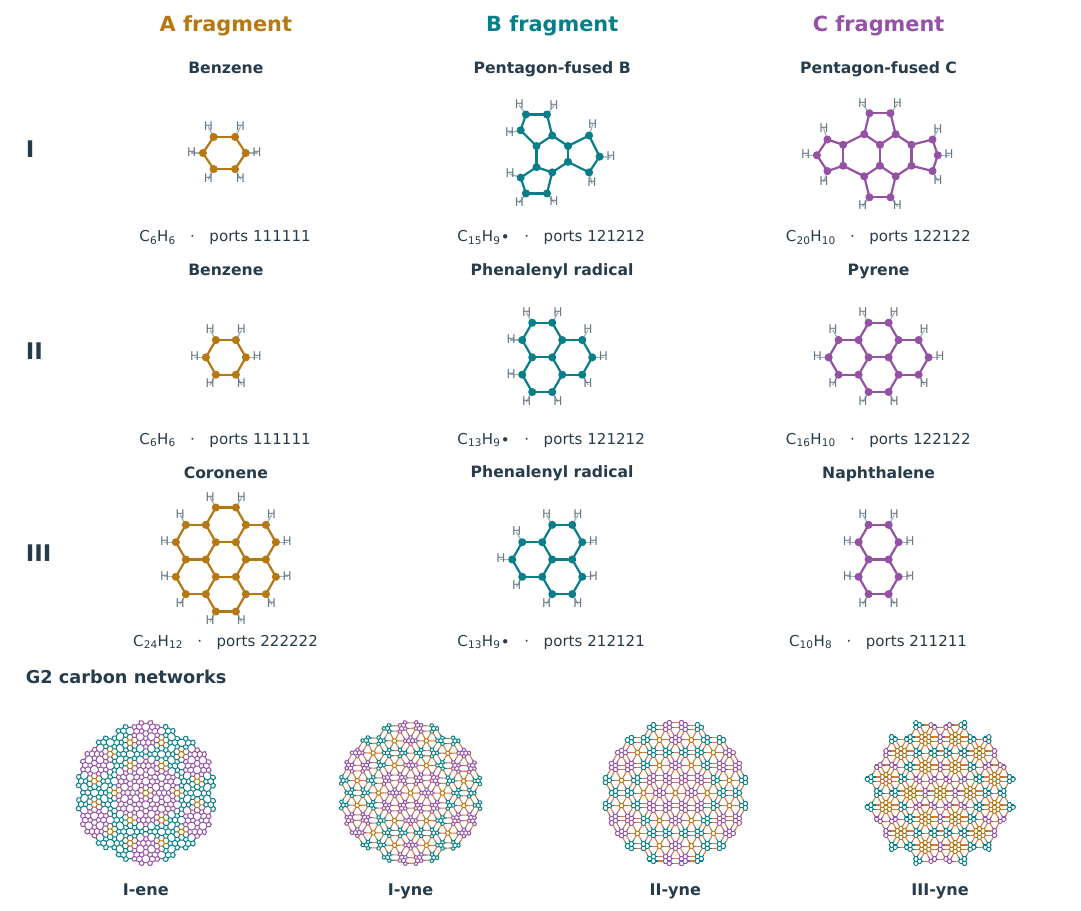}
\caption{\textbf{Three molecular alphabets and four candidate networks.} The H-capped fragment diagrams identify each carbon skeleton. The six digits below each fragment give the port multiplicities around its perimeter. H caps are replaced at assembly connections. The lower row shows the four G2 networks, with core colours inherited from A/B/C. Direct assembly of alphabets II and III yields graphene and is therefore not counted as a distinct candidate. The drawings show the construction geometries.}\label{fig:chemistry}
\end{figure}

Each letter carries an oriented carbon fragment and six directional ports. A port contains one or two terminal atoms; matching its multiplicity and orientation across every triangular-lattice edge determines the assembly. Figure~\ref{fig:chemistry} shows the three alphabets. Hydrogen caps describe isolated fragment boundaries and are removed where fragments connect. The periodic sheets discussed here contain carbon only.

Alphabet I uses a six-carbon benzene core, a 15-carbon hexagon decorated by three pentagons, and a 20-carbon core comprising two fused hexagons with four pentagons. Their H-capped formulas are C$_6$H$_6$, C$_{15}$H$_9$ and C$_{20}$H$_{10}$. Alphabet II uses benzene, phenalenyl and pyrene, with formulas C$_6$H$_6$, C$_{13}$H$_9$ and C$_{16}$H$_{10}$. Alphabet III uses coronene, phenalenyl and naphthalene, C$_{24}$H$_{12}$, C$_{13}$H$_9$ and C$_{10}$H$_8$, with complemented port multiplicities. An odd number of hydrogens in these neutral fragments signals an odd electron count; the capped drawings are not an assertion that every fragment is an isolated closed-shell synthetic precursor. They define candidate carbon skeletons and connection sites, not an established synthesis route.

We consider two link constructions. In an ene network the matched terminal carbons are bonded directly. In a yne network two additional carbon atoms are inserted per terminal pair, producing an ethynylene connection, C--C$\equiv$C--C. The latter are graphyne-type networks; a graphdiyne link would instead contain four inserted carbons. These labels define connectivity, while the bond lengths and out-of-plane response remain relaxation variables.

Only four distinct candidates are retained. Directly joining the all-benzenoid fragments of alphabets II and III recovers the ordinary honeycomb network and erases the distinction between molecular blocks. Explicit assembly of A1B0C3, A4B6C5, A7B8C12 and A13B18C18 for each alphabet reduces to a two-carbon graphene primitive cell (SI). Ethynylene insertion retains the marked connections, producing II-yne and III-yne. Alphabet I has nonhexagonal connectivity and gives both I-ene and I-yne. For a closed periodic network with molecular counts $(A,B,C)$, the carbon counts are
\begin{equation}\label{eq:carbon}
\begin{aligned}
\Ncarbon^{\mathrm{I-ene}}&=6A+15B+20C,&\quad
\Ncarbon^{\mathrm{I-yne}}&=12A+24B+30C,\\
\Ncarbon^{\mathrm{II-yne}}&=12A+22B+26C,&
\Ncarbon^{\mathrm{III-yne}}&=36A+22B+18C.
\end{aligned}
\end{equation}
Finite samples use explicit atom counts because open ports change the linker contribution. All four generation series supplied here use the same complete-ABC supports, allowing their hierarchy to be compared without changing the molecular boundary convention.

\section{Periodic basis construction and an independent holdout}
Periodic calculations enter through two routes. The first assigns oriented A/B/C units on a periodic triangular supercell and tests every port and neighbourhood across its periodic boundaries. The second extracts a cell from the \emph{already coloured} G3 structure, identifies opposite edges, and then rechecks the periodic seams. It does not recolour a new hexagon. A cut can introduce environments absent from the infinite field, so extraction alone does not certify an admissible energy basis.

We retain only cells whose first-shell environments are a subset of the seven in Eq.~\eqref{eq:f}. In the archived set of 123 records, seven pass this criterion; after accounting for translated duplicates and repetitions of smaller patterns, the relevant patterns are A1B0C3, A4B6C5, A7B8C12 and A13B18C18. For example, the archived $N=48$ cut fails because its seam creates A:BBBBBC and C:ABBABB centres. It is not included in the basis or in the energy validation.

\begin{figure}[p]\centering
\includegraphics[width=\linewidth]{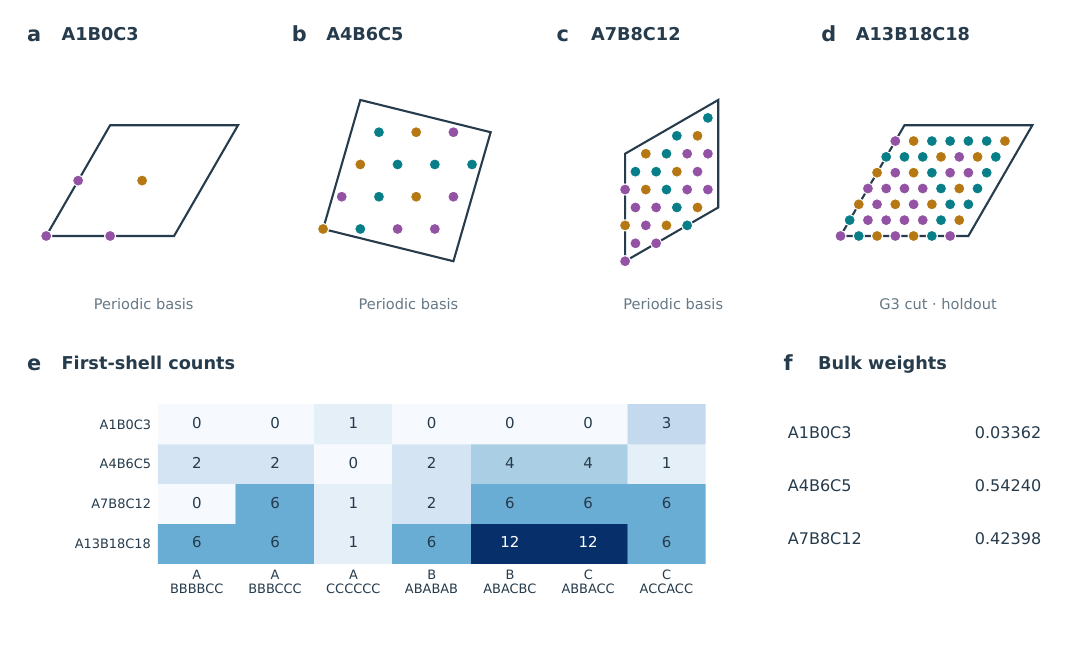}
\caption{\textbf{An exact local-statistics basis.} The three periodic basis cells and the G3-derived A13B18C18 holdout are shown on the triangular lattice. The count matrix uses the same seven labels as Fig.~\ref{fig:hierarchy}. The displayed normalized weights match molecular-site frequencies. Cell total energies must instead be combined with Eq.~\eqref{eq:energy}, including its atom-count denominator.}\label{fig:basis}
\end{figure}

Figure~\ref{fig:basis} gives the environment count vectors of the selected cells. The basis vectors have rank three and satisfy the exact identity
\begin{equation}\label{eq:basis}
(23\beta+7)\mathbf f_\infty
=(4-\beta)\mathbf n_{A1B0C3}+3\mathbf n_{A4B6C5}+(\beta-2)\mathbf n_{A7B8C12}.
\end{equation}
When each vector is divided by its number of molecular sites, the corresponding weights are $0.0336158194$, $0.5424049825$ and $0.4239791982$. These are molecular-site weights, not atom weights or weights for cell total energies.

Consider the first-shell local-energy approximation for a fixed chemical family,
\begin{equation}\label{eq:local}
E_j=\sum_{\mu=1}^{7}n_{j\mu}\epsilon_\mu+\mathcal R_j.
\end{equation}
The residual $\mathcal R_j$ includes longer-range electronic contributions, elastic relaxation and any local information omitted by the seven labels. Three basis energies cannot determine seven independent $\epsilon_\mu$. Nevertheless, Eq.~\eqref{eq:basis} determines the target linear functional uniquely within this model:
\begin{equation}\label{eq:energy}
\widehat e_\infty=
\frac{(4-\beta)E_{A1B0C3}+3E_{A4B6C5}+(\beta-2)E_{A7B8C12}}
{(4-\beta)N_{\mathrm C,A1B0C3}+3N_{\mathrm C,A4B6C5}+(\beta-2)N_{\mathrm C,A7B8C12}}.
\end{equation}
The hats emphasize an energy estimate. Exact matching of counts does not set $\mathcal R_j$ to zero, and each chemical family requires its own three energies.

The A13B18C18 cell is held out from the basis construction and obeys
\begin{equation}\label{eq:holdout}
\mathbf n_{A13B18C18}=\mathbf n_{A1B0C3}+3\mathbf n_{A4B6C5},\qquad
\widehat E_{A13B18C18}=E_{A1B0C3}+3E_{A4B6C5}.
\end{equation}
Its actual DFT energy gives a directly interpretable validation residual
$\delta_{A13B18C18}=(E_{A13B18C18}-\widehat E_{A13B18C18})/N_{\mathrm C,A13B18C18}$.
This tests transfer across one different periodic arrangement; it does not independently probe the A7B8C12 direction or prove the accuracy of the infinite-limit estimate. The SI details this limitation and how to extend the validation if necessary.

\section{Energy transfer from periodic structures to the bronze limit}
The four families give bulk estimates of $-9.037523$ eV/C for I-ene and $-8.661320$, $-8.661665$ and $-8.684749$ eV/C for I-, II- and III-yne, respectively (Table~\ref{tab:summary}). III-yne lies about 23 meV/C below the other two ethynylene networks on this scale. I-yne and II-yne differ by only 0.345 meV/C, less than the errors in their periodic transfer tests.

\begin{figure}[p]\centering
\includegraphics[width=\linewidth]{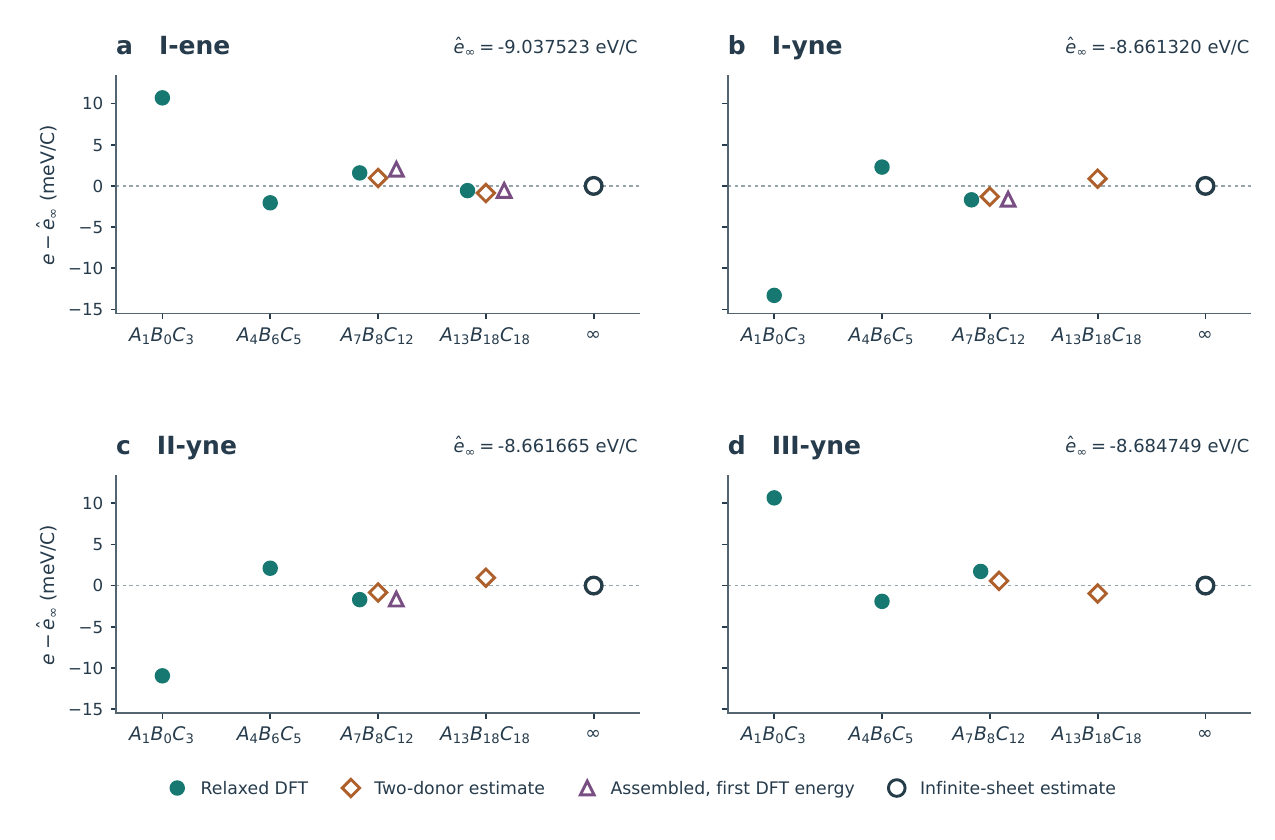}
\caption{\textbf{Periodic cell energies and the infinite-sheet estimate.} In each panel the energy per carbon is measured relative to the infinite-sheet estimate $\hat e_\infty$ for that family. The open circle at $\infty$ marks this estimate, whose absolute value is given above the panel. Filled circles show the original relaxed DFT energies, open triangles the first DFT energies of the assembled cells, and diamonds the two-donor predictions. The A7B8C12 prediction uses projected environment counts. The A13B18C18 prediction uses the exact count identity in Eq.~\eqref{eq:holdout}. Symbols at the same composition are offset horizontally for clarity.}\label{fig:energy}
\end{figure}

We tested the energy model at two levels (Fig.~\ref{fig:energy} and Fig.~S14). For the 708-carbon A13B18C18-ene-I cell, the exact count identity in Eq.~\eqref{eq:holdout} gives $-6399.179034$ eV, compared with $-6398.970251$ eV from DFT, a residual of $+0.295$ meV/C. Two smaller I-ene cells with exact count identities in the extended first-shell catalogue give $-0.121$ and $-0.219$ meV/C. A7B8C12 provides a more demanding test. Its seven environment counts span a third independent direction. Before using its relaxed energy in the bulk estimate, we predicted its energy from A1B0C3 and A4B6C5 alone, by projecting its environment frequencies onto the line between the two donors (SI). The four signed residuals run from $-0.866$ to $+1.147$ meV/C. These comparisons test a specific transfer prescription and do not bound the infinite-system error.

We also tested the transfer of atomic positions in four periodic cells with 402--708 carbons, covering two directly bonded and two ethynylene-linked structures. VASP relaxations starting from the assembled geometries lower the first reported total energy by 0.016--0.042 meV/C. The root-mean-square atomic shifts are 0.00061--0.00241~\AA\ after accounting for changes in the cell. The root-mean-square changes in actual C--C bond lengths, including cell deformation, are 0.00070--0.00113~\AA. All four retain their complete periodic bond graphs (Fig.~S17). Figure~\ref{fig:energy} shows the first DFT energy of each assembled cell. The corresponding relaxation traces are given in Fig.~S17. The relaxed basis energies used in the linear model retain their original DFT references.

\begin{table}[htbp]
\centering\small
\caption{\textbf{Bulk estimates and periodic energy tests for four carbon networks.} The first block uses the exact A13B18C18 count identity. The second uses the energy-independent two-donor projection for A7B8C12. $\widehat e_\infty$ and $e_{\rm DFT}$ are in eV/C, $\widehat E$ in eV/cell and $\delta=(E_{\rm DFT}-\widehat E)/N_{\rm C}$ in meV/C. A dash denotes a large-cell DFT calculation not performed.}\label{tab:summary}
\resizebox{\linewidth}{!}{\begin{tabular}{@{}l r rrr rrr@{}}
\toprule
 & & \multicolumn{3}{c}{A$_{13}$B$_{18}$C$_{18}$: exact counts} & \multicolumn{3}{c}{A$_7$B$_8$C$_{12}$: projected counts}\\
\cmidrule(lr){3-5}\cmidrule(lr){6-8}
Family & $\widehat e_\infty$ & $\widehat E$ & $e_{\rm DFT}$ & $\delta$ & $\widehat E$ & $e_{\rm DFT}$ & $\delta$\\
\midrule
I-ene & -9.037523 & -6399.1790 & -9.038094 & +0.295 & \multicolumn{3}{c}{See Fig.~S14}\\
I-yne & -8.661320 & -9768.9785 & -- & -- & -5509.4289 & -8.663010 & -0.386\\
II-yne & -8.661665 & -8833.9241 & -- & -- & -4954.9536 & -8.663373 & -0.866\\
III-yne & -8.684749 & -10318.6311 & -- & -- & -5592.6117 & -8.683033 & +1.147\\
\bottomrule
\end{tabular}
}
\end{table}
For the three A13B18C18 yne sheets, Table~\ref{tab:summary} reports predictions from the exact count combination. Their electronic spectra use geometries assembled from relaxed donors. The energy estimates characterize relative energetics within this series. An assessment of dynamical or finite-temperature stability would require evidence beyond these total energies and local relaxation tests.

\section{Periodic bands and states of open carbon patches}
We used the distance-dependent, nonorthogonal carbon $p_z$ model of gt-TB\cite{Gong2020,He2021} for the periodic bands, total $\pi$ density of states (DOS) and finite-fragment levels. Thirteen periodic geometries are DFT-relaxed. The other three are the assembled A13B18C18 yne geometries. For each cell, the band path and the Brillouin-zone DOS share one energy zero. We integrate over a two-dimensional triangular mesh without Gaussian broadening. On the combined path and mesh, A13B18C18-ene-I has a 654.2 meV separation and the III-yne analogue has a 113.6 meV separation. I-yne shows a 17.4 meV indirect overlap. For II-yne the mesh separation falls from 19.8 to 2.4 meV on refinement, leaving its classification sensitive to sampling. Figures S8--S11 and the SI sampling table identify these differences for all sixteen cells. The one-orbital model probes the out-of-plane $\pi$ sector.

All four G3 fragments use the same 943-site sequence, $A_{235}B_{324}C_{384}$. We fit the carbon coordinates in one $60^\circ$ sector to local distances from the three relaxed periodic basis cells of the same family. The other five sectors follow by rotation, with bonds across their boundaries included throughout the fit (Fig.~S16). The resulting planar coordinates recover every prescribed bond. The finite-fragment spectra below use these sector-fitted coordinates.

\begin{figure}[p]\centering
\includegraphics[width=.96\linewidth,height=.76\textheight,keepaspectratio]{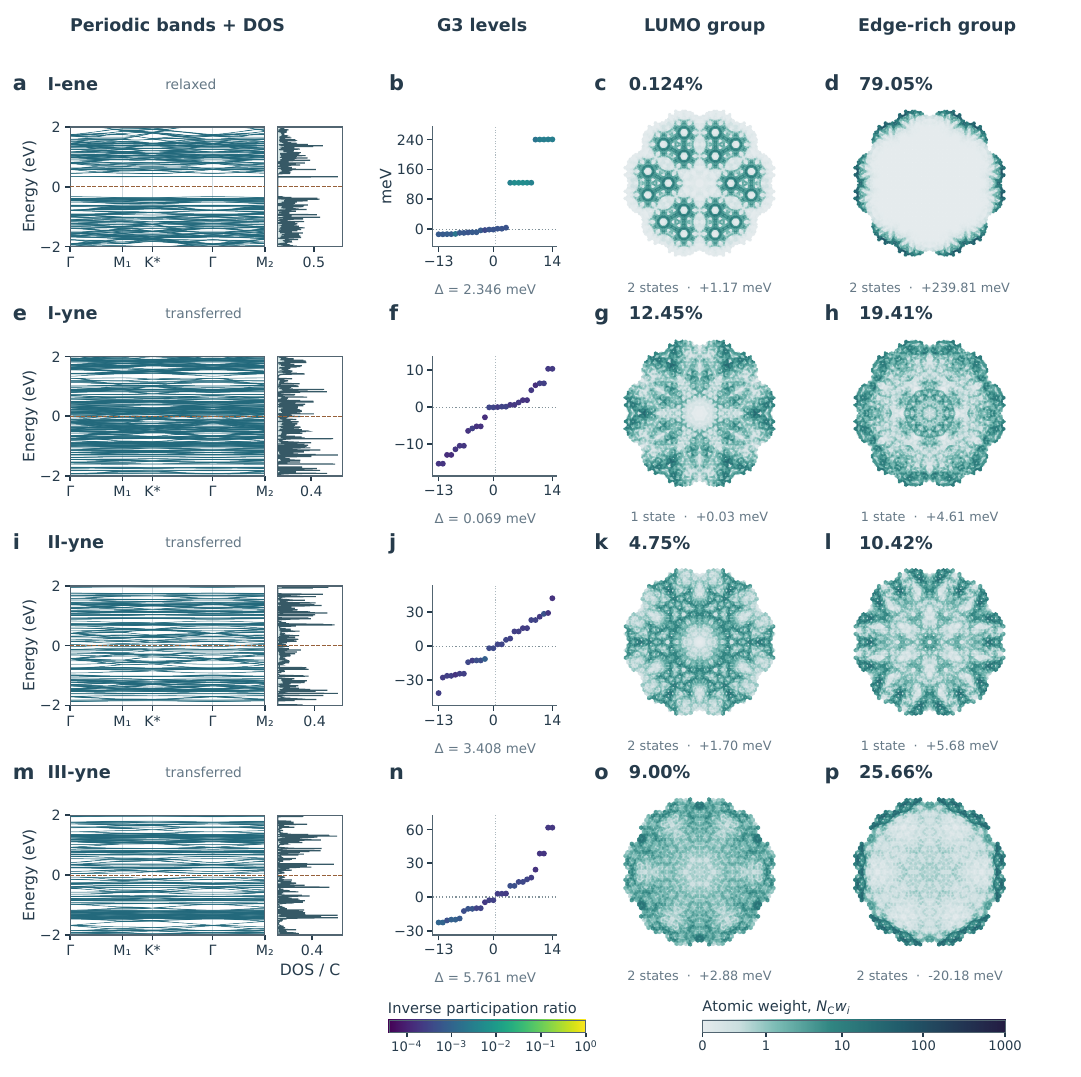}
\caption{\textbf{Periodic spectra and boundary weights in the sixfold G3 fragments.} Rows show the four chemical families. The first column gives A13B18C18 $\pi$ bands and spin-summed DOS in states per eV per carbon. The second plots 28 G3 energies relative to the neutral midpoint, against orbital index relative to the HOMO (index 0). Colour gives the inverse participation ratio of each solver eigenvector and $\Delta$ is the HOMO--LUMO spacing. The last two columns show the LUMO group and the complete interior-window group with the largest boundary weight on the same fragment. Percentages are populations within two bonds of an undercoordinated carbon. The energies below the maps use the fragment midpoint. The colour scale gives the mean atomic population divided by the uniform value $1/N_{\rm C}$.}\label{fig:electronic}
\end{figure}

The four complete G3 fragments have HOMO--LUMO spacings of 2.346, 0.069, 3.408 and 5.761 meV for I-ene, I-yne, II-yne and III-yne, respectively (Fig.~\ref{fig:electronic}). We define the boundary shell as all atoms within two carbon bonds of an undercoordinated atom. Among the complete energy groups in the 28-level frontier window, excluding its endpoint groups, the highest shell weights are 79.05\%, 19.41\%, 10.42\% and 25.66\%, in the same order. The corresponding shells contain 7.96\%, 6.14\%, 6.22\% and 5.80\% of the carbon atoms.

I-ene separates boundary enrichment from the assignment of the LUMO. Its LUMO doublet has only 0.124\% of its population in the shell. A higher doublet at $+239.81$ meV from the neutral HOMO--LUMO midpoint carries the 79.05\% shell weight and extends along the six sides. On the same carbon graph, a fit without the rotation constraint gives a LUMO localized mainly on one edge, with 72.33\% shell weight. The root-mean-square change in atomic positions between the fits is only 0.0176~\AA. A second effect concerns how a doublet is displayed. An individual real eigenvector can give unequal weights in the six sectors even for the exactly symmetric structure. The population averaged over the complete LUMO doublet recovers sixfold symmetry (Fig.~S18). These results concern the carbon-only termination and the specified $\pi$ Hamiltonian.

\section{Discussion}
The most restrictive choice appears at the scale of three neighbouring sites. A B triangle can be coloured quasiperiodically as part of an ordered infinite field, yet it prevents the alternating carbon ports from closing. The bronze hierarchy supplies the compatible seven-environment catalogue for the carbon networks studied here. Its local catalogue has two roles: it constrains which molecular assemblies are permitted, and it defines an explicit energy approximation that can be challenged by a held-out calculation. The exact count identity avoids fitting an underdetermined set of seven environment energies; it does not remove the need to test electronic and elastic nonlocality.

The four selected networks isolate two chemical choices: nonhexagonal versus benzenoid cores, and direct versus ethynylene connections. The reduction of II-ene and III-ene to graphene is itself instructive: an aperiodic partition of identical atoms does not ensure an aperiodic observable network. The decoration must retain a physical record of the code. Establishing that record in relaxed geometry, and identifying an electronic response that depends on its hierarchy, are the decisive next tests. The periodic basis also supplies local distances for finite fragments. Building the rotation into that fit makes it possible to distinguish the geometry sensitivity of a frontier orbital from the basis freedom within a degenerate pair.

\section*{Methods and data availability}
We preoptimized the original carbon-network geometries with RG$^2$\cite{Shi2018RG2}, then relaxed the periodic reference structures in VASP\cite{Kresse1996}. In the geometry-transfer tests, the starting positions came from the relaxed donors of the same chemical family. The SI gives the VASP settings recorded for these tests and the gt-TB parameters used in the spectral calculations\cite{Gong2020,He2021}. The accompanying files include the integer constructions and certificates, indexed carbon structures, cell energies and relaxation traces, periodic eigenvalues, DOS integration and open-patch eigenstate populations.
\clearpage

\clearpage
\clearpage
\setcounter{section}{0}
\renewcommand{\theHsection}{supp.\arabic{section}}
\renewcommand{\thesection}{S\arabic{section}}
\setcounter{subsection}{0}
\renewcommand{\theHsubsection}{supp.\arabic{section}.\arabic{subsection}}
\setcounter{figure}{0}
\renewcommand{\theHfigure}{supp.\arabic{figure}}
\renewcommand{\thefigure}{S\arabic{figure}}
\setcounter{table}{0}
\renewcommand{\theHtable}{supp.\arabic{table}}
\renewcommand{\thetable}{S\arabic{table}}
\setcounter{equation}{0}
\renewcommand{\theHequation}{supp.\arabic{equation}}
\renewcommand{\theequation}{S\arabic{equation}}

\begin{center}{\Large\bfseries Supplementary Information\\\large From quasiperiodic three-colour order on a triangular lattice to covalent carbon sheets\par}\medskip\manuscriptauthor\end{center}
\noindent We give the colouring and local-port proofs, the periodic energy basis, independent energy and geometry tests, and the electronic calculations for four carbon networks. The periodic spectra use thirteen relaxed and three transferred structures. The finite G3 spectra use sector fits to the complete ABC support.

\section*{Supplementary contents}
\noindent\hyperref[combined-si-sec-1]{S1\quad Golden, silver and bronze colour fields}\dotfill\pageref{combined-si-sec-1}\par
\noindent\hyperref[combined-si-sec-2]{S2\quad Definitions, scale parameters and finite supports}\dotfill\pageref{combined-si-sec-2}\par
\noindent\hyperref[combined-si-sec-3]{S3\quad Axis certificate and absence of planar translations}\dotfill\pageref{combined-si-sec-3}\par
\noindent\hyperref[combined-si-sec-4]{S4\quad Growing collars and recurrence of arbitrary finite patches}\dotfill\pageref{combined-si-sec-4}\par
\noindent\hyperref[combined-si-sec-5]{S5\quad Molecular alphabets, ports and carbon counting}\dotfill\pageref{combined-si-sec-5}\par
\noindent\hyperref[combined-si-sec-6]{S6\quad Why II-ene and III-ene are not retained}\dotfill\pageref{combined-si-sec-6}\par
\noindent\hyperref[combined-si-sec-7]{S7\quad Periodic-cell generation and seam eligibility}\dotfill\pageref{combined-si-sec-7}\par
\noindent\hyperref[combined-si-sec-8]{S8\quad Exact basis algebra and identifiability}\dotfill\pageref{combined-si-sec-8}\par
\noindent\hyperref[combined-si-sec-9]{S9\quad Periodic energy checks and infinite-limit estimates}\dotfill\pageref{combined-si-sec-9}\par
\noindent\hyperref[combined-si-sec-10]{S10\quad Electronic calculations and structural validation}\dotfill\pageref{combined-si-sec-10}\par
\noindent\hyperref[combined-si-sec-11]{S11\quad Data and reproduction}\dotfill\pageref{combined-si-sec-11}\par
\noindent\hyperref[combined-si-sec-12]{S12\quad Supplementary figures}\dotfill\pageref{combined-si-sec-12}\par

\section{Golden, silver and bronze colour fields}\label{combined-si-sec-1}
Use integer coordinates $(i,j)$ on 60-degree axes, and take the norm $\rho=\max(|i|,|j|,|i+j|)$. Write $H(r)=\{\rho\leq r\}$. The common rule in the main text uses $m=1,2,3$, with $(p,q)\mapsto(q,mq+p)$. For the gold and silver branches all sites are eventually candidates. Gold coverage follows by induction. The base contains $H(2)$ after two steps. If the current patch contains $H(p)$ and the next pair is $(p,q)$, take $x\ge y\ge0$ by symmetry and $p<x+y\le q$. Subtract $(q-p,0)$. Since $q\le2p$, the result is in $H(p)$. The axial copies and the original patch cover $H(q)$.

For silver, four steps cover $H(28)$ and the next pair is $(29,70)$. A three-step certificate propagates $H(0.9p)$ to $H(0.9(2p+5q))$. Normalize $p=1$. The ratio remains in $[70/29,169/70]$ under $t\mapsto2+1/t$. The certificate enumerates 2,893 possible translation coefficient quadruples and 64,516 grid witnesses in one symmetry sector at spacing $1/40$. The maximum distance to a legal translated centre on that grid is at most $224/280$. Adding the gridding error $2/40$ and the parameter error $11(169/70-70/29)$ gives $3473/4060<0.9$. Subtracting the chosen centre from any integer target therefore gives an integer point in the smaller patch. The exact integer checks are in \path{code/metallic/certify_candidates.py}. The following sections give the separate bronze certificate.

For gold and silver the A condition reduces to
\[
\max(\|id\|,\|jd\|,\|(i+j)d\|)<d/2 ,
\]
where $\|\cdot\|$ denotes distance to an integer, with $d=1/\varphi^2$ and $\sqrt2-1$, respectively. The phase pair $(id,jd)\bmod 1$ samples a two-torus uniformly. The intersection of the three selecting strips has area $3d^2/4$, the irrational A density. A period would give an independent period under a 60-degree rotation and force rational density. The orbit misses all window boundaries since $d$ is irrational. Each finite patch therefore occupies an open set in phase space and recurs with bounded gaps by compactness. Integer scales $(p_h,q_h)$ yield small phase errors and repeated finite patches. The exact raw-copy recursion and the recurrence of the final gold and silver colour fields are distinct statements.

A site's six-neighbour word is determined by a radius-two A patch. Partitioning the phase torus along all translated boundaries of that patch gives the complete finite list and exact areas in \path{code/metallic/new_environments.csv} (Fig.~S15). Each seven-site wheel has twelve internal edges. Enumerating all rotations of the three port sequences checks all twelve edges simultaneously. Complementing every port as in alphabet III preserves equality of matching ports. For the stated assignment only five of eight golden wheels and six of nine silver wheels pass. Global decoration is excluded by any failed wheel. The B triangle in the main text gives an explicit obstruction; it occurs inside the all-B wheel. Trying all six global reassignments of the colours to the motifs yields a failing wheel in every case (\texttt{ports\_all\_assignments.csv}).
\section{Definitions, scale parameters and finite supports}\label{combined-si-sec-2}
We distinguish the integer recursion level $h$, the finite carbon-support label G1/G2/G3, and the periodic-cell index. Periodic structures are named by their molecular composition, linkage and alphabet, for example A13B18C18-ene-I. The holdout contains 49 molecular sites. The precise site lists, orientations and matrices are retained in the accompanying JSON files.

The bronze number $\beta=(3+\sqrt{13})/2$ is the asymptotic recursion ratio. The geometric length ratio $(\sqrt3+\sqrt{39})/6\simeq1.32951$ associated with the parent bronze construction\cite{Dotera2017} is a different quantity. A chemically chosen ratio, such as the earlier ethynylene embedding value near 1.65425, is also not the bronze inflation factor. Integer colouring, parent geometric embedding and relaxed chemical distances must be reported separately.

The common finite supports used for all four candidates are listed in Table~\ref{tab:gen}. They are complete molecular-block supports extracted from the common archived hierarchy. A raw finite set $S_h$ of A digits is not itself the G$h$ carbon molecule. For reproducibility, the present generation convention is defined by the supplied site lists; changes to boundary pruning constitute a different finite sample, even if the same interior colour field is used.
\begin{table}[htbp]\centering
\caption{Common finite supports and explicit carbon counts before optional H capping.}\label{tab:gen}
\begin{tabular}{lrrrrrrr}\toprule
Support & A & B & C & I-ene & I-yne & II-yne & III-yne\\\midrule
G1 & 1 & 0 & 6 & 126 & 162 & 138 & 120\\
G2 & 19 & 36 & 30 & 1254 & 1902 & 1710 & 1908\\
G3 & 235 & 324 & 384 & 13950 & 21774 & 19590 & 22140\\\bottomrule
\end{tabular}
\end{table}
Earlier I-only files used a differently truncated atomic boundary and had carbon counts 72/1104/13296 (ene) and 108/1740/21072 (yne). These should not be mixed with the matched-support series in Table~\ref{tab:gen}. All four G3 calculations here use the complete support of Table~\ref{tab:gen} and its original carbon connectivity. Optional H-capped structures are controls for termination effects, not alternative bulk compositions.

\section{Field and statements}
Let $U=\{(1,0),(0,1),(-1,1),(-1,0),(0,-1),(1,-1)\}$, in cyclic order.
Put $\rho(i,j)=\max(|i|,|j|,|i+j|)$ and
\[
 p_0=1,\quad q_0=3,\qquad p_{h+1}=q_h,\quad q_{h+1}=3q_h+p_h.
\]
With indices on $U$ understood modulo six, define
\[
 D_h=\{0\}\cup\{(q_h-p_h)u_k\}
       \cup\{q_hu_k+p_hu_{k\pm1}\},\qquad
 S_0=\{0\},\quad S_{h+1}=S_h+D_h,\quad S_\infty=\bigcup_h S_h.
\]
The colour of $x\in\mathbb Z^2$ is A when $x\in S_\infty$;
otherwise it is C when $x-u,x+u\in S_\infty$ for at least one $u\in U$,
and B otherwise. Thus A takes priority. All sums are sets, with duplicates
removed.

A first-shell environment records the centre and the cyclic word of its six
neighbours, up to rotations and reflections. This is distinct from a pair
interface, which records only the colours at the ends of one edge.

\begin{theorem}
The field has exactly the seven first-shell environments in
Table~\ref{tab:freq}. Their frequencies, defined by counting centres in
expanding hexagonal lattice windows, are the tabulated values. The same
limits hold in translated large hexagonal windows and ordinary planar
windows whose boundary contribution vanishes relative to their area.
Consequently, the molecular-site fractions are
\[
 (f_A,f_B,f_C)
 =\frac{1}{4+\sqrt{13}}\bigl(2,\sqrt{13}-1,3\bigr).
\]
\end{theorem}

\begin{table}[ht]
\centering
\caption{Exact frequencies. Here $\beta=(3+\sqrt{13})/2$ and
$D=23\beta+7$. The denominator is the total number of molecular lattice
sites, not the number of atoms.}\label{tab:freq}
\begin{tabular}{lll}
\toprule
Environment & Exact frequency & Percentage\\\midrule
A:BBBBCC & $6/D$ & $7.23206643$\\
A:BBBCCC & $6(\beta-1)/D$ & $16.65382639$\\
A:CCCCCC & $2/D$ & $2.41068881$\\
B:ABABAB & $2(\beta+1)/D$ & $10.37265309$\\
B:ABACBC & $6\beta/D$ & $23.88589282$\\
C:ABBACC & $6\beta/D$ & $23.88589282$\\
C:ACCACC & $3(\beta+1)/D$ & $15.55897963$\\\bottomrule
\end{tabular}
\end{table}

\section{Finite access to the infinite field}
Let $R_h=\sum_{k<h}(p_k+q_k)$. Every sum of digits below level $h$ has norm
at most $R_h$, and induction gives $R_h<2p_h$. A sum whose highest nonzero
digit is at level $h$ has norm at least
\[
 H_h=q_h-p_h-R_h,\qquad H_0=2,\qquad H_{h+1}=H_h+p_h.
\]
These inequalities are integer bounds and do not assume uniqueness of
digit expansions.

For the square in integer coordinates
$W_h=[0,q_h-p_h)^2\cap\mathbb Z^2$, all A sites needed to decode the
first shells have norm at most $2(q_h-p_h)$. The square is a
parallelogram in the triangular lattice. Higher digits beyond $h+2$
cannot contribute, since
\[
 H_{h+3}=5q_h+p_h-R_h>5q_h-p_h>2(q_h-p_h).
\]
Hence a three-level zero-origin guard contains every required high digit.

\section{Rectangular subdivision and the collar}
Use three interval lengths
\[
 L_X(h)=q_h-p_h,\qquad L_Y(h)=3p_h,\qquad L_Z(h)=q_h+2p_h.
\]
The integer identities give the substitutions
\[
 X\longrightarrow XYX,\qquad
 Y\longrightarrow XZX,\qquad
 Z\longrightarrow XZXZX.
\]
Take Cartesian products of these interval subdivisions. Their purpose here
is to partition the lattice into rectangles in integer coordinates; no
assumption that the two-dimensional colour is a product of one-dimensional
words is made.

A state is $(T_x,T_y,\mathcal S)$, where each
$(a,b,c,d)\in\mathcal S$ denotes a processed high-digit sum
$(aq_h+bp_h,cq_h+dp_h)$ relative to the rectangle origin. The digit
coefficient set is
\[
 \mathcal D=\{(0,0,0,0)\}
 \cup\{(x,-x,y,-y):(x,y)\in U\}
 \cup\{(x,v,y,w):(v,w)\text{ adjacent to }(x,y)\text{ in }U\}.
\]
On descending one level, with child offsets
$Q_xq_h+P_xp_h$, $Q_yq_h+P_yp_h$, update by
\[
 (a,b,c,d)\mapsto
 (3a+b+e-Q_x,\ a+f-P_x,\
  3c+d+g-Q_y,\ c+k-P_y)
\]
for every $(e,f,g,k)\in\mathcal D$.

An environment at a centre requires A membership out to two lattice rings:
one ring to access its neighbours and another for the C midpoint test.
Unprocessed lower digits and these two rings require at most
$R_h+2<2p_h+2\le4p_h$ in each coordinate. Therefore retain only a
conservative $4p_h$ collar around the current rectangle.

This retention test is performed exactly for
$r=q_h/p_h\in[3,10/3]$. If $L_T(h)/p_h=u r+v$, discard a coordinate
$ar+b$ only if it is below $-4$ at both endpoints, or if
$(a-u)r+(b-v)$ exceeds $4$ at both endpoints. A linear function cannot
cross the retained interval without satisfying at least one of these
endpoint checks. The test may keep unnecessary digits, but cannot discard
a digit needed by the rectangle. Every retained digit remains an actual
digit sum, so extra retention cannot introduce spurious A sites.

Starting from zero and applying three zero-offset X--X guard steps,
breadth-first closure under all child rectangles gives 301 states:
\[
 \begin{array}{c|rrr}
  T_x\backslash T_y&X&Y&Z\\\hline
  X&81&42&36\\Y&42&21&16\\Z&36&16&11
 \end{array}.
\]
The accompanying JSON lists every quadruple of every state and every
transition. The verifier reconstructs all of them from the equations.
At $h=0$ the side lengths are $2,3,5$ and each quadruple becomes
$(3a+b,3c+d)$. Decoding each base rectangle produces exclusively the
seven environments in Table~\ref{tab:freq}.

By induction, closure is valid at every depth. The growing root squares
cover the nonnegative coordinate sector, and the sixfold symmetry of the
digits and colour test covers the entire plane. Thus there is no eighth
first-shell environment. Every listed environment occurs in a decoded
state reachable from the root, so the list contains exactly seven types.

\section{Frequencies and exact algebra}
Let $M_{ij}$ count children of state $i$ of type $j$, and let
$O_{i\mu}$ count environment $\mu$ in the base rectangle of state $i$.
Then $M^h O$ gives all seven counts at depth $h$.
The state graph is strongly connected, and 16 states have self-loops.
Thus the nonnegative matrix $M$ is primitive.

Write
$\ell_X=\beta-1$, $\ell_Y=3$, $\ell_Z=\beta+2$.
The positive vector $v_i=\ell_{T_x}\ell_{T_y}$ satisfies
$Mv=\beta^2v$ by the product interval subdivision. The Perron
eigenvalue is therefore $\lambda=\beta^2=(11+3\sqrt{13})/2$.
Primitivity gives a common environment-frequency limit for all states.

For a compact exact computation of that limit, the verifier checks the
following integer identity on all $301\times7$ output entries:
\[
 M^3(M^2-11M+I)(M^2-3M-I)(M^2-I)\,O=0.
\]
No fitted recurrence or rounded eigenvector is used. With
$\lambda_-=(11-3\sqrt{13})/2$, set
\[
 Q(x)=x^3(x-\lambda_-)(x^2-3x-1)(x^2-1).
\]
All roots of the displayed annihilating polynomial other than $\lambda$
have modulus smaller than $\lambda$. Applying $Q(M)$ isolates the
dominant contribution to the seven output vectors. Its root-state row,
normalized by its sum, simplifies exactly to
\[
 \frac{1}{23\beta+7}
 \bigl(6,\,6(\beta-1),\,2,\,2(\beta+1),\,6\beta,\,6\beta,\,
       3(\beta+1)\bigr).
\]
The supplied SymPy checks verify these identities in $\mathbb Q(\sqrt{13})$.

To transfer this result from superrectangles to large windows, fix a
subdivision level $h$. All level-$h$ rectangle diameters are bounded by
a constant times $q_h$. Partition a sufficiently large root square down
to this level. Rectangles meeting the boundary of a large hexagon
contribute only a strip of this fixed width, whose fraction vanishes as
the hexagon grows. Every interior rectangle has, uniformly over the
finite state set, the same limiting frequency vector as $h$ grows.
First increase the window size, then increase $h$. Apply the same
argument sector by sector; sector-boundary strips also have vanishing
density. This establishes the stated planar frequencies, including
translated windows.

\section{Pair interfaces and finite samples}
There are five nonzero undirected colour-pair interfaces:
AB, AC, BB, BC and CC. AA is absent. If $f_1,\ldots,f_7$ follow
Table~\ref{tab:freq}, counting each edge once and using $3N$ edges
for $N$ bulk triangular-lattice sites gives
\[
\begin{aligned}
 g_{AB}&=(4f_1+3f_2)/3=f_A,&
 g_{AC}&=(2f_1+3f_2+6f_3)/3=f_A,\\
 g_{BB}&=(3f_4+2f_5)/6=f_A/2,&
 g_{BC}&=2f_5/3,\\
 g_{CC}&=(2f_6+4f_7)/6.
\end{aligned}
\]
Their percentages are respectively
$26.2966$, $26.2966$, $13.1483$, $15.9239$,
and $18.3346$. These identities are independent of whether a
particular chemical family resolves additional port multiplicities or
orientation classes; those extra chemical distinctions require their
own labels.

The common complete-ABC G3 support contains $235$ A, $324$ B and
$384$ C units. Its finite fractions differ from the bulk limits.
Direct finite-window statistics in the package count all centres with
their full neighbours supplied by a halo, avoiding artificial
open-boundary environments. The radius-6000 window has
$108018001$ sites and yields $(0.2629509039,0.3425589037,0.3944901924)$,
consistent with the exact limit but not used to prove it.
Direct and recursively counted root rectangles were also checked
independently through side length $2738$, with exact agreement.

All four retained chemical networks sharing this colouring have the same
molecular-site frequencies. Atomic fractions, relaxed energies,
termination counts, and graph changes caused by reconstruction are
separate quantities.

\section{Axis certificate and absence of planar translations}\label{combined-si-sec-3}
The interval lengths $L_X=q-p$, $L_Y=3p$, $L_Z=q+2p$ obey
$X\mapsto XYX$, $Y\mapsto XZX$, $Z\mapsto XZXZX$ on descending one level. The one-dimensional restriction is verified with the accompanying nine-state exact axis program. Its states retain the transverse A sources required to decide whether a site is C; consequently, this is not an arbitrary one-dimensional rule imposed on a two-dimensional pattern. At the base level, every state of type X, Y or Z decodes respectively to AC, ABB or ABBBB.

With columns denoting the substituted parent, the interval incidence matrix is
\begin{equation}
M_1=\begin{pmatrix}2&2&3\\1&0&0\\0&1&2\end{pmatrix}.
\end{equation}
It is primitive with Perron eigenvalue $\beta$. If the interval densities are proportional to $(x,y,z)$, then $x=\beta y$ and $y=(\beta-2)z$. Since each base word has one A and lengths 2, 3, 5, the axis A density is
\begin{equation}
\frac{x+y+z}{2x+3y+5z}=\frac{2\beta}{5\beta+1}.
\end{equation}
This is irrational: a rational value $r$ would imply $(5r-2)\beta=-r$, impossible for irrational $\beta$. A periodic word has a rational letter density, so the axis restriction is not periodic.

The full colour field is invariant under $60^\circ$ rotation $R(a,b)=(-b,a+b)$, because its digit sets and local midpoint rule are invariant. If a nonzero $t=(a,b)$ were a translation period, so would $Rt$ and every integer combination of them. In particular,
\begin{equation}
(a+b)t-bRt=(a^2+ab+b^2,0)
\end{equation}
would be a nonzero horizontal period, contradicting the axis result. Thus the field admits no nonzero planar translation. This argument excludes even a single translation period, rather than merely excluding a two-dimensional lattice of periods.

\section{Growing collars and recurrence of arbitrary finite patches}\label{combined-si-sec-4}
Primitivity of a local counting matrix by itself would not prove recurrence of arbitrarily large patches. Here the rectangle states retain more information: the same retained A sources determine all colours in a collar whose physical width grows with level.

At level $h$, put $\delta_h=2p_h-1$. Consider a state rectangle enlarged by $\delta_h$ in each integer coordinate. To colour any point in this enlarged rectangle requires its own A membership and that of its nearest neighbours. The maximum lower-digit displacement is $R_h<2p_h$, and
\begin{equation}\label{eq:collar}
R_h+\delta_h+1<4p_h.
\end{equation}
Therefore the conservative $4p_h$ source collar used by the certificate suffices to determine the entire enlarged coloured rectangle. During a descent, the child rectangle lies in its parent and $\delta_h$ decreases; a source pruned at an ancestor cannot affect this smaller target. This proves the growing-collar property by induction.

The three root guard steps also suffice for this larger target. The A sources needed for the enlarged X--X rectangle have norm bounded by $2(q_h-p_h)+2\delta_h+2$, whereas
\begin{equation}
H_{h+3}=5q_h+p_h-R_h>5q_h-p_h>
2(q_h-p_h)+2\delta_h+2.
\end{equation}
Here $q_h\ge3p_h$. Consequently a root state at level $h$ contains the same origin-centred coloured ball of radius $\delta_h$ as the infinite field, and $\delta_h\to\infty$.

The verified transition graph is strongly connected, the root has a self-loop, and every state reaches the root in at most three descents. Padding with the root self-loop gives a root descendant after exactly three descents for every state. These finite statements are checked in \texttt{code/verify\_recurrence.py}. Fix any finite patch of the original field and choose an origin-centred ball containing it. Choose $h$ so that this ball lies within the root's $\delta_h$ collar. Every level-$(h+3)$ state rectangle then contains a root copy at level $h$ and hence a translated copy of the patch. Rectangle sizes at this fixed level are bounded, so the gaps between copies are bounded.

Growing root squares cover the nonnegative coordinate sector; sixfold rotations cover the plane. The origin-centred colour ball is itself sixfold invariant, so rotated root copies contain translated copies of the same ball. Sector boundaries do not create unbounded gaps. Thus every finite coloured patch recurs relatively densely: the field is uniformly recurrent and aperiodic.

As an additional finite check, all 301 states were compared against an independently generated field throughout collars of widths 1, 5 and 19 at levels 0, 1 and 2, with 1,556,082 cumulative point comparisons and no mismatch. This numerical check corroborates the induction but does not replace it. The graph and frequency proofs are reproducible from integer data supplied with this manuscript.

\paragraph{What is and is not proved.} The results establish a well-defined, aperiodic and uniformly recurrent real-space hierarchy with exact local frequencies. We use ``quasiperiodic code'' in this structural sense. Pure-point dynamical or diffraction spectrum is a stronger mathematical property\cite{Solomyak1997,Solomyak1999}. No theorem asserting it for the decorated or relaxed atomic sheets is supplied here. A finite diffraction image, if later added, should be labelled a finite-size diagnostic rather than a proof of that spectral property.

\section{Molecular alphabets, ports and carbon counting}\label{combined-si-sec-5}
Table~\ref{tab:ports} records the fragments before assembly. Port strings are read cyclically in the archived local frame; rotation changes their displayed starting direction. The B cores have odd neutral electron counts when capped as shown. Calling them fragments avoids implying that all capped motifs have been synthesized or are stable closed-shell reagents.
\begin{table}[htbp]\centering\small
\caption{Three molecular alphabets. $n$ is the number of core carbon atoms; $P$ is the number of external terminal valences.}\label{tab:ports}
\begin{tabular}{lllrll}\toprule
Alphabet & Letter & Capped fragment & $n$ & Port string & $P$\\\midrule
I & A & Benzene, C$_6$H$_6$ & 6 & 111111 & 6\\
I & B & Pentagon-fused, C$_{15}$H$_9$ & 15 & 121212 & 9\\
I & C & Pentagon-fused, C$_{20}$H$_{10}$ & 20 & 122122 & 10\\
II & A & Benzene, C$_6$H$_6$ & 6 & 111111 & 6\\
II & B & Phenalenyl, C$_{13}$H$_9$ & 13 & 121212 & 9\\
II & C & Pyrene, C$_{16}$H$_{10}$ & 16 & 122122 & 10\\
III & A & Coronene, C$_{24}$H$_{12}$ & 24 & 222222 & 12\\
III & B & Phenalenyl, C$_{13}$H$_9$ & 13 & 212121 & 9\\
III & C & Naphthalene, C$_{10}$H$_8$ & 10 & 211211 & 8\\\bottomrule
\end{tabular}
\end{table}

Let $L$ count matched terminal pairs. A periodic ene cell has $N_{\rm C}=\sum_s n_s N_s$, and a yne cell has $N_{\rm C}=\sum_s n_sN_s+2L$. In a fully closed periodic network, $2L=\sum_s P_sN_s$, so the yne coefficients are simply $n_s+P_s$. This derives the four carbon-count formulas in the main text. It also shows why one cannot count only coarse triangular edges: some edges carry two atomic connections. For an open finite patch, $L$ must be counted explicitly and any remaining valences treated by a declared termination convention.

In I-yne the limiting carbon atoms per molecular site equal $16+2\sqrt{13}$; for I-ene they equal $11+\sqrt{13}$. For II-yne and III-yne they are respectively
$(80+22\sqrt{13})/(4+\sqrt{13})$ and $(104+22\sqrt{13})/(4+\sqrt{13})$.
These are atom-count normalizations, not independent concentration variables.

\section{Why II-ene and III-ene are not retained}\label{combined-si-sec-6}
The direct benzenoid assemblies were reconstructed on all four selected periodic supports. At the unrelaxed ideal geometry, enumeration of the translations permuting the carbon sites with a Cartesian tolerance of $10^{-5}$ \AA\ gives two carbon atoms for every cell (Table~\ref{tab:graphene}). The reconstructed bond-vector residuals are below $10^{-8}$ \AA. These checks establish that the selected direct assemblies are graphene supercells: their A/B/C boundaries are a partition of the honeycomb network, not a distinct carbon topology. The molecular code becomes physically visible again when the connecting bonds are replaced with ethynylene links. Figure S4 illustrates the difference.
\begin{table}[htbp]\centering
\caption{Direct-assembly checks. All eight ideal structures reduce to a two-carbon graphene primitive cell. These are geometric checks, not DFT relaxations.}\label{tab:graphene}
\begin{tabular}{lrrrr}\toprule
Assembly & A1B0C3 & A4B6C5 & A7B8C12 & A13B18C18\\\midrule
II-ene carbon count & 54 & 182 & 338 & 600\\
III-ene carbon count & 54 & 224 & 392 & 726\\
Primitive carbon count, both & 2 & 2 & 2 & 2\\\bottomrule
\end{tabular}
\end{table}

\section{Periodic-cell generation and seam eligibility}\label{combined-si-sec-7}
For the periodic-colouring route, a nonsingular integer matrix $T$ defines a triangular supercell with $N=|\det T|$ molecular sites. Periodic representatives are assigned letters and orientations; neighbours are found modulo $T$. A candidate must satisfy all port matches, geometric assembly checks and the chosen local-environment criterion. The archived catalogue contains 118 accepted construction records from the earlier small-supercell search, plus five records cut from G3. The present work does not claim that this 123-record archive enumerates all possible cells or all possible larger-period colourings.

For the cut route, retain the letters and orientations already present in G3 and wrap the selected window by its two translation vectors. Recompute every first shell using periodic neighbours, including sites on the seam. The accepted A13B18C18 window contains 49 sites and begins at $(-9,-2)$ in integer coordinates; a parallelogram fundamental domain can represent the same hexagonal Bravais superlattice. The illustration in Fig. S7 shows this fundamental domain rather than implying a recolouring operation. Source site lists and matrices are authoritative for reproduction.

Eligibility for the seven-environment energy model requires
\begin{equation}
\mathcal E_j\subseteq\mathcal E_\infty,\qquad \mathcal E_\infty=\{\mu_1,\ldots,\mu_7\},
\end{equation}
where the seven labels are listed in Table~\ref{tab:freq}. Having fewer than seven types is allowed. A periodic environment outside this set is disqualifying for the present basis, even if the atomic assembly is chemically admissible. The seven passing records comprise the four compositions used in the main text, translated copies of A1B0C3 and A7B8C12, and a threefold repetition of A1B0C3 with composition A3B0C9. These duplicates do not increase the descriptor rank. A13B20C15 is excluded because it contains six A:BBBBBC and three C:ABBABB centres.

\section{Exact basis algebra and identifiability}\label{combined-si-sec-8}
With the environment order used in the main text, the count matrix is
\begin{equation}\label{eq:counts}
X=\begin{pmatrix}
0&0&1&0&0&0&3\\
2&2&0&2&4&4&1\\
0&6&1&2&6&6&6\\
6&6&1&6&12&12&6
\end{pmatrix},
\end{equation}
whose rows are A1B0C3, A4B6C5, A7B8C12 and A13B18C18. Its rank is three. The first three rows contain 4, 15 and 27 molecular sites; the last contains 49. Put $D=23\beta+7$. Direct substitution gives
\begin{equation}
D\mathbf f_\infty=(4-\beta)X_1+3X_2+(\beta-2)X_3,
\qquad X_4=X_1+3X_2.
\end{equation}
The nonnegative molecular-site weights are
\begin{equation}
\mathbf w=\frac{(4(4-\beta),45,27(\beta-2))}{D},\qquad\sum_jw_j=1.
\end{equation}
For an energy per carbon atom, the corresponding weights of the three \emph{per-carbon} energies would be
\begin{equation}
\widetilde w_j=\frac{a_j N_{{\rm C},j}}{\sum_k a_kN_{{\rm C},k}},\qquad
\mathbf a=(4-\beta,3,\beta-2).
\end{equation}
They depend on the chemical family. Using $\mathbf w$ directly on per-carbon energies generally gives the wrong result.

An alternative expression is $E=\sum_sN_s\eta_s+\sum_\mu n_\mu J_\mu$ for molecular and interface contributions. Since $N_s=\sum_{\mu:\,\mathrm{centre}(\mu)=s}n_\mu$, the ten parameters are redundant. They can be absorbed into seven effective $\epsilon_\mu=\eta_{\mathrm{centre}(\mu)}+J_\mu$, which are still underdetermined by three basis energies. No unique set of seven interface energies should be reported from these inputs. The target is nevertheless identifiable because its descriptor lies in the row span of the basis.

If $E_j=X_j\boldsymbol\epsilon+\mathcal R_j$, the error in the infinite per-carbon estimate is
\begin{equation}
\widehat e_\infty-e_\infty
=\frac{\sum_{j=1}^{3}a_j\mathcal R_j}{D\nu_\infty}
-\frac{r_\infty}{\nu_\infty},
\end{equation}
where $\nu_\infty$ is the carbon count per molecular site and $r_\infty$ the missing residual energy per molecular site in the infinite structure. Exact frequencies alone bound neither term. The held-out residual measures
\begin{equation}
E_4-E_1-3E_2=\mathcal R_4-\mathcal R_1-3\mathcal R_2.
\end{equation}
It does not test $\mathcal R_3$ independently. Thus a small holdout error supports a limited transfer test, not a rigorous error bar on $e_\infty$.

A second algebraic estimate can use A4B6C5, A7B8C12 and A13B18C18 with normalized molecular weights $(0.1642270145,0.4239791982,0.4117937874)$. Its difference from the primary estimate is proportional to the same holdout residual, so the two estimates are not independent validations. Further validation should use genuinely different longer-range arrangements while preserving the allowed local environments, or direct finite-patch energy calculations with controlled boundary corrections. Merely repeating a small periodic cell adds no independent information.

For a finite patch, $E_G$ generally includes boundary and termination terms. A schematic extension is
\begin{equation}
E_G\simeq\sum_\mu n_{G\mu}\epsilon_\mu+\sum_b N_{G,b}\gamma_b+\mathcal R_G.
\end{equation}
The boundary coefficients require independent determination. The infinite-limit basis formula should not be applied directly to H-capped G3 total energies without accounting for hydrogen and boundary chemistry.

\section{Periodic energy checks and infinite-limit estimates}\label{combined-si-sec-9}
\begin{table}[htbp]\centering\small
\caption{Relaxed periodic reference energies. A dash denotes a large-cell DFT calculation not performed. Relaxations initialized from the donor geometries are documented separately below.}\label{tab:energies}
\begin{tabular}{lrrr}
\toprule Structure & $N_{\rm C}$ & $E$ (eV/cell) & $e$ (eV/C)\\\midrule
A1B0C3-ene-I & 66 & -595.770492 & -9.026825636\\
A4B6C5-ene-I & 214 & -1934.469514 & -9.039577168\\
A7B8C12-ene-I & 402 & -3632.449374 & -9.035943716\\
A13B18C18-ene-I & 708 & -6398.970251 & -9.038093575\\
A1B0C3-yne-I & 102 & -884.811397 & -8.674621539\\
A4B6C5-yne-I & 342 & -2961.389040 & -8.659032281\\
A7B8C12-yne-I & 636 & -5509.674498 & -8.663010217\\
A13B18C18-yne-I & 1128 & -- & --\\
A1B0C3-yne-II & 90 & -780.536587 & -8.672628744\\
A4B6C5-yne-II & 310 & -2684.462510 & -8.659556484\\
A7B8C12-yne-II & 572 & -4955.449142 & -8.663372626\\
A13B18C18-yne-II & 1020 & -- & --\\
A1B0C3-yne-III & 90 & -780.668458 & -8.674093978\\
A4B6C5-yne-III & 366 & -3179.320884 & -8.686669082\\
A7B8C12-yne-III & 644 & -5591.873110 & -8.683032780\\
A13B18C18-yne-III & 1188 & -- & --\\
\bottomrule
\end{tabular}
\end{table}
The twelve relaxed basis energies enter the infinite-limit estimate for their respective chemistries. A13B18C18 is outside this fit. Its predicted total energy and carbon count are exact combinations of the same two donors in all four families, subject to the local-energy assumption for the energy. The complete cell totals are
\begin{center}\small\begin{tabular}{lrrr}
\toprule
Family & C atoms & Predicted (eV/cell) & Relaxed DFT (eV/cell)\\
\midrule
I-ene & 708 & -6399.179034 & -6398.970251\\
I-yne & 1128 & -9768.978517 & --\\
II-yne & 1020 & -8833.924117 & --\\
III-yne & 1188 & -10318.631110 & --\\
\bottomrule
\end{tabular}
\end{center}
The two auxiliary I-ene checks obey the first-shell count identities
\[
 {\bf n}_{\rm A3B2C7}={\bf n}_{\rm A1B0C3}+{\bf n}_{\rm A2B2C4},\qquad
 {\bf n}_{\rm A6B10C6}={\bf n}_{\rm A2B4C1}+{\bf n}_{\rm A4B6C5}.
\]
The two auxiliary donor energies are $-1102.169062$ eV (122 carbons) and $-832.238499$ eV (92 carbons) for A2B2C4 and A2B4C1, respectively. These checks count every environment on the periodic cell, including A:BBCCCC in the first identity and A:BBBBBC and C:ABBABB in the second. Only the seven bronze environments enter the infinite-limit calculation.

\paragraph{Testing A7B8C12 with its energy withheld.}
Let $j=0,1,t$ denote A1B0C3, A4B6C5 and A7B8C12, let $M=(4,15,27)$ be their molecular site counts and set $f_j={\bf n}_j/M_j$. The three count vectors have rank three. We define an approximate two-donor estimate by minimizing the unweighted squared difference between seven site frequencies, with no energy input from the target or any other target in this series:
\[
 \lambda=\frac{(f_t-f_0)\cdot(f_1-f_0)}{\|f_1-f_0\|^2}=0.78884632,
\]
\[
 u=\frac{M_t(1-\lambda)}{M_0}=1.42528736,\qquad
 v=\frac{M_t\lambda}{M_1}=1.41992337 .
\]
For each family we use the same $u,v$ and compute
\[
 \widehat E_t=N_{{\rm C},t}\,
 \frac{uE_0+vE_1}{uN_{{\rm C},0}+vN_{{\rm C},1}} .
\]
The denominator accounts for the different carbon contents of the decorations and preserves shifts of the energy zero per carbon. This projection approximates the environment vector. For example, A:BBBBCC occurs zero times in A7B8C12 but has positive weight in the donor mixture. The test measures both the approximation in the descriptor and energy transfer. After the test, the measured $E_t$ supplies the independent third basis energy used for the bulk estimate.
\begin{table}[htbp]\centering\small\caption{Signed transfer errors. Exact-count tests and projected tests use only donor energies in the prediction. DFT references are the original relaxed totals. The signed error is DFT minus predicted energy per carbon.}\label{tab:energytests}
\begin{tabular}{llrrr}
\toprule
Cell & Count match & $\widehat E$ (eV) & $E_{\rm DFT}$ (eV) & $\delta$ (meV/C)\\
\midrule
A3B2C7-ene-I & exact & -1697.939554 & -1697.962322 & -0.121\\
A6B10C6-ene-I & exact & -2766.708013 & -2766.775138 & -0.219\\
A13B18C18-ene-I & exact & -6399.179034 & -6398.970251 & +0.295\\
A7B8C12-ene-I & projection & -3632.698237 & -3632.449374 & +0.619\\
A7B8C12-yne-I & projection & -5509.428870 & -5509.674498 & -0.386\\
A7B8C12-yne-II & projection & -4954.953630 & -4955.449142 & -0.866\\
A7B8C12-yne-III & projection & -5592.611684 & -5591.873110 & +1.147\\
\bottomrule
\end{tabular}
\end{table}

\section{Electronic calculations and structural validation}\label{combined-si-sec-10}
\subsection{Periodic DFT and local geometry}
We preoptimized the original periodic frameworks with RG$^2$\cite{Shi2018RG2} before relaxation in VASP\cite{Kresse1996}. For the three A7B8C12 transfer tests, the supplied VASP 6.4.3 outputs identify PAW--PBE carbon with four valence electrons\cite{Blochl1994,PBE1996}. These checks used a 400 eV cutoff, $\Gamma$ sampling, Gaussian electronic smearing of 0.1 eV, an electronic tolerance of $10^{-5}$ eV and an ionic force tolerance of $0.02$ eV/\AA. The conjugate-gradient ionic relaxations allowed cell-shape changes at fixed volume (ISIF=4). Each run reached the specified stopping condition. For A13B18C18-ene-I, the retained evidence comprises the assembled coordinates, the relaxed RES export and eight reported ionic total energies. We use the archived periodic reference energies consistently for the bulk estimate. The calculations characterize relaxation on the sampled energy surfaces. We report neither phonon nor finite-temperature data.

\subsection{Local geometry transfer and open boundaries}
The large periodic starting structures use the relaxed A1B0C3 and A4B6C5 geometries of the same family. The open fragments use all three relaxed basis structures. We label each donor bond and each pair of neighbours of a common central atom by the A/B/C owner and the $sp$ or $sp^2$ class. Each half of an ethynylene link is assigned to the adjacent core. Successive descriptors collect the sorted neighbour signatures over up to three graph shells. A target takes the deepest matching descriptor. Among donors bearing that descriptor, we retain those whose ideal reference-length mismatch is within 0.014~\AA\ of the best match and average their relaxed distances. This gives a target $d_t^*$ for each indexed pair $t$. The neighbour-pair distances constrain the angles between adjacent bonds. G1 and G2 use planar fits over all carbon atoms.

Figure~\ref{fig:assembly} shows the construction and its four carbon realizations. For complete G3, the lattice action $(i,j)\mapsto(-j,i+j)$ fixes the central A and permutes the remaining molecular sites in orbits of six. The open carbon graph supplies a $60^\circ$ permutation $P$ preserving the molecule assignments, carbon types, connectivity and edge mask. Each sector contributes 39 A, 54 B and 64 C sites in addition to the shared central A. Each carbon orbit, including those in the central molecule, has six members. Select a representative in $0\leq\theta<60^\circ$ about the fragment centroid. For carbon $i$, let $a(i)$ index its representative and let $k(i)$ give its sector relative to that representative. With $R$ the planar $60^\circ$ rotation and $y_a$ the independent carbon coordinates, fit
\begin{equation}\label{eq:sector-fit}
x_i=R^{k(i)}y_{a(i)},\qquad
\min_{\{y_a\}}\sum_{t=(i,j)}\left(\lVert x_i-x_j\rVert-d_t^*\right)^2 .
\end{equation}
The sum includes crossing bonds and crossing neighbour pairs. Six rotated copies of an indexed pair have one fitted distance. Averaging their targets and weighting by their multiplicity changes the objective only by a coordinate-independent constant. The $y_a$ can move both radially and tangentially throughout the fit. We choose the free overall azimuth by alignment with the construction geometry and export coordinates in the original atom order. The sixfold fit uses the complete ABC template as the starting geometry. The open carbon-only fragments have no added hydrogens.

\begin{table}[htbp]\centering\small
\caption{Geometry of the fitted G3 sectors. $N_{\rm sec}$ counts independent carbon atoms and $B_{\rm seam}$ counts crossing bonds over all six boundaries. Lengths are deviations from the donor bond targets. The last column gives the maximum absolute crossing-bond deviation.}\label{tab:seams}
\begin{tabular}{lrrrrr}\toprule
Family & $N_{\rm sec}$ & $B_{\rm seam}$ & Inner RMS (\AA) & Seam RMS (\AA) & Seam max. (\AA)\\\midrule
I-ene & 2,325 & 294 & 0.01037 & 0.02085 & 0.06839\\
I-yne & 3,629 & 306 & 0.00517 & 0.01035 & 0.02719\\
II-yne & 3,265 & 282 & 0.00104 & 0.00051 & 0.00307\\
III-yne & 3,690 & 288 & 0.00068 & 0.00042 & 0.00181\\\bottomrule
\end{tabular}\end{table}

For all four exported G3 structures, the $1.85$~\AA\ neighbour search recovers the full input bond graph. The maximum displacement between corresponding carbons under a $60^\circ$ rotation is below $2.6\times10^{-12}$~\AA, and the maximum difference in corresponding bond lengths is below $2.2\times10^{-12}$~\AA. Table~\ref{tab:seams} and Fig.~\ref{fig:residuals} quantify deviations from donor lengths independently of these symmetry checks. The G3 structures have only undergone geometric fitting.

\paragraph{Four paired geometric checks.}
For each of the three A7B8C12 cases we compare the POSCAR of the transferred start to the CONTCAR of the \emph{same} relaxation. The coordinates of A13B18C18-ene-I are available as the two archived geometries. Its final lattice parameters and fractional coordinates are rounded to three and six decimal places, respectively. For atom $i$ we minimum-image its difference in fractional coordinates, map that vector into the final in-plane cell, include its vertical change, and subtract the mean displacement. We report the RMS and maximum norm of this internal shift. For each indexed C--C bond $b$, the signed change in length is $d_b^{\rm last}-d_b^{\rm start}$, using the actual periodic lattice at each endpoint. The bond RMS includes cell deformation. The area change is $100(A_{\rm last}/A_{\rm start}-1)$. All intended bonds survive in each case at a 1.85~\AA\ cutoff. The 13 original reference coordinates likewise retain their intended periodic graphs. The G3 metric compares final fitted bond lengths to their donor targets, with the internal and cross-sector bonds counted separately.
\begin{table}[htbp]\centering\small
\caption{Changes during relaxation from transferred geometries. Maxima are absolute values. The energy drop is the difference between first and last reported total energy per carbon of the same run.}\label{tab:paired}
\resizebox{\linewidth}{!}{\begin{tabular}{lrrrrrrr}
\toprule
 & & \multicolumn{2}{c}{Atomic shift (\AA)} & \multicolumn{2}{c}{Bond change (\AA)} & Area & Energy drop\\
\cmidrule(lr){3-4}\cmidrule(lr){5-6}
Cell & $N_{\rm C}$ & RMS & Max. & RMS & Max. & (\%) & (meV/C)\\
\midrule
A7B8C12-ene-I & 402 & 0.00241 & 0.00524 & 0.00075 & 0.00201 & +0.0273 & 0.0370\\
A7B8C12-yne-I & 636 & 0.00225 & 0.00633 & 0.00113 & 0.00324 & +0.0064 & 0.0419\\
A7B8C12-yne-II & 572 & 0.00061 & 0.00213 & 0.00088 & 0.00402 & +0.0039 & 0.0157\\
A13B18C18-ene-I & 708 & 0.00126 & 0.00429 & 0.00070 & 0.00272 & +0.0644 & 0.0201\\
\bottomrule
\end{tabular}
}\end{table}
\begin{table}[htbp]\centering\small
\caption{First and last DFT energies of the four relaxations initialized from assembled coordinates. The rightmost column gives the independent reference used to evaluate the energy combinations.}\label{tab:pairedE}
\begin{tabular}{lrrrr}
\toprule
Cell & Entries & First (eV) & Last (eV) & Original reference (eV)\\
\midrule
A7B8C12-ene-I & 10 & -3632.272060 & -3632.286929 & -3632.449374\\
A7B8C12-yne-I & 14 & -5509.632005 & -5509.658662 & -5509.674498\\
A7B8C12-yne-II & 5 & -4955.414198 & -4955.423160 & -4955.449142\\
A13B18C18-ene-I & 8 & -6398.955997 & -6398.970251 & -6398.970251\\
\bottomrule
\end{tabular}
\end{table}
For A7B8C12-ene-I, -yne-I and -yne-II respectively, the final energies of the three restarted structures exceed the corresponding original relaxed references by 0.404, 0.025 and 0.045 meV/C. Table~\ref{tab:pairedE} lists the two final energies side by side. Figure~\ref{fig:residuals} reports each relaxation trace relative to its own last entry. Our energy-combination tests consistently use the original relaxed reference totals. The raw traces and the paired geometries permit either comparison to be reproduced.

\subsection{Carbon \texorpdfstring{$p_z$}{pz} Hamiltonian, bands and DOS}
We used the five-parameter TB20 implementation of the multi-neighbour gt-TB carbon model\cite{Gong2020,He2021} and the specified 1.426~\AA\ reference length. For distinct sites within 10~\AA,
\[
H_{ij}=-3.02\exp[2.11(1-r_{ij}/1.426)]~\mathrm{eV},\qquad
S_{ij}=0.235\exp[2.01(1-r_{ij}/1.426)] .
\]
Take $H_{ii}=0$ and $S_{ii}=1$ for the same site and image. Periodic-image terms carry the Bloch phase $\exp(2\pi i\,\mathbf k_{\rm frac}\cdot\mathbf T)$ in the cell gauge. We solved $Hc=ESc$. Each carbon supplies one $p_z$ electron with spin degeneracy two. The result describes the out-of-plane $\pi$ sector, including on acetylenic carbons. Spin splitting, $\sigma$ states, in-plane $\pi$ orbitals on $sp$ carbons and changes to boundary on-site energies are outside this model.

Bloch bands use 11 points per segment with shared endpoints counted once. In the reciprocal basis of each reported cell the path visits $\Gamma$, $\mathbf b_1/2$, the adjacent Wigner--Seitz vertex common to the bisectors of $\mathbf b_1$ and $\mathbf b_1+\mathbf b_2$, then $\Gamma$ and $\mathbf b_2/2$. The vertex is labelled V for the oblique cells and K* for near-hexagonal metrics. These labels specify sampled geometrical points. The two-dimensional DOS uses periodic, unshifted $36^2$, $24^2$, $12^2$ and $12^2$ meshes for A1B0C3, A4B6C5, A7B8C12 and A13B18C18, respectively. We split each reciprocal parallelogram along its $(1,1)$ diagonal, interpolate every sorted band linearly within each triangle, and integrate its occupied area analytically between successive energy-bin edges. The nominal width is 2 meV. Bins are split at the sampled gap edges and at the common midgap zero for positive separations. The DOS is plotted as a bin average without Gaussian broadening or gap masking. Its integral is two states per carbon with spin. For a positive separation the energy zero is the midpoint between the highest occupied and lowest unoccupied energy over the union of mesh and path. For a band overlap both plots share the median eigenvalue of the uniformly sampled mesh. A signed separation compares the extrema of bands $N_{\rm C}/2$ and $N_{\rm C}/2+1$. The nested half-density mesh is a sampling diagnostic. In particular, a small positive separation on the sampled points does not prove a gap throughout the Brillouin zone.
\begin{table}[htbp]\centering\small
\caption{Sampling of the frontier bands. Separations in meV are evaluated over the indicated uniform mesh, the displayed path, or their union. The DOS integrates the $n\times n$ mesh. A negative value denotes overlap. Values close to zero should be interpreted at the mesh resolution.}\label{tab:meshes}
\begin{tabular}{lrrrrr}
\toprule
Cell & $n$ & Mesh $n/2$ & Mesh $n$ & Path & Union\\
\midrule
A1B0C3-ene-I & 36 & +459.27 & +459.27 & +459.27 & +459.27\\
A4B6C5-ene-I & 24 & +739.34 & +738.53 & +740.25 & +738.53\\
A7B8C12-ene-I & 12 & +620.27 & +620.27 & +620.27 & +620.27\\
A13B18C18-ene-I & 12 & +654.19 & +654.19 & +654.19 & +654.19\\
A1B0C3-yne-I & 36 & -230.09 & -252.56 & -239.39 & -252.56\\
A4B6C5-yne-I & 24 & -47.46 & -50.22 & -33.85 & -50.22\\
A7B8C12-yne-I & 12 & +4.97 & +4.23 & +4.04 & +4.04\\
A13B18C18-yne-I & 12 & -5.06 & -12.22 & -17.44 & -17.44\\
A1B0C3-yne-II & 36 & +40.39 & +16.33 & +35.37 & +16.33\\
A4B6C5-yne-II & 24 & +35.60 & +35.60 & +40.36 & +35.60\\
A7B8C12-yne-II & 12 & +0.05 & -0.61 & -0.61 & -0.61\\
A13B18C18-yne-II & 12 & +19.76 & +2.44 & +2.44 & +2.44\\
A1B0C3-yne-III & 36 & +315.59 & +315.59 & +315.59 & +315.59\\
A4B6C5-yne-III & 24 & +0.17 & +0.17 & +11.25 & +0.17\\
A7B8C12-yne-III & 12 & +0.03 & +0.03 & +0.03 & +0.03\\
A13B18C18-yne-III & 12 & +113.59 & +113.59 & +113.59 & +113.59\\
\bottomrule
\end{tabular}
\end{table}

\subsection{Finite levels and boundary weights}
All finite spectra use open boundaries. G1 and G2 use their full-fragment fits and G3 uses the sector fit of Eq.~\eqref{eq:sector-fit}. For each neutral patch the reported 28 levels comprise 14 occupied and 14 unoccupied levels about its HOMO--LUMO midpoint. For G3 a sparse shift-invert solver targets the frontier. Negative pivots of the symmetrically permuted real $H-\mu S$ locate the half-filling index. The implementation checks the row and column permutations, residuals and positive overlap. For $S$-normalized eigenvectors the positive atomic populations are $w_i=|(S^{1/2}c)_i|^2$. The inverse participation ratio is $\sum_i w_i^2$ for each normalized solver eigenvector. Within a degenerate subspace this single-vector quantity can depend on the basis. Orbital maps below use the complete group unless an individual vector is indicated.

Boundary seeds are atoms with fewer neighbours than their bulk $sp$ or $sp^2$ coordination. The mask includes all atoms within two graph bonds of a seed. We also archive the results for one and three bonds. To select a map, adjacent eigenvalues separated by at most $10^{-5}$ eV are grouped; the mean atomic weight of the group gives a basis-independent population for a degenerate subspace. We select the complete group with the highest average boundary fraction in the reported window, omitting groups touching its endpoints. Selection yields an upper envelope over the tested states. Our measures establish spatial concentration at the model boundary. They carry no assertion of protected edge or corner modes.

\subsection{Geometry sensitivity and the population of a degenerate level}
For each sector-fitted G3, every pair inside the 10~\AA\ hopping cutoff maps to another included pair under $P$. With $U$ the permutation matrix, the relative Frobenius errors $\|U^{\mathsf T} H U-H\|_{\rm F}/\|H\|_{\rm F}$ and $\|U^{\mathsf T} S U-S\|_{\rm F}/\|S\|_{\rm F}$ are below $9\times10^{-13}$ and $4\times10^{-13}$, respectively. We compare these sector-fit spectra with a separate fit of all atomic coordinates without a symmetry constraint, using the same complete ABC support, atom order, donor targets, open edges and Hamiltonian parameters. The electronic problem is solved on each set of coordinates.

For a group of $g$ states the plotted weight is $w_i=g^{-1}\sum_n |(S^{1/2}c_n)_i|^2$. Its rotational mismatch is
\begin{equation}
D_{60}=\tfrac12\sum_i |w_i-w_{P(i)}|.
\end{equation}
For a complete degenerate eigenspace the population is invariant under a change of basis. Its spectral projector commutes with the rotation when $H$ and $S$ do. An individual real eigenvector in a doublet can give a direction-dependent density. The $10^{-5}$~eV grouping threshold also specifies the resolution used to compare levels on the unconstrained structures.

\begin{table}[htbp]\centering\small
\caption{LUMO symmetry on the same indexed carbon graph before and after imposing the sector constraint. The RMS displacement compares the two geometries after rigid alignment. The next two columns give the LUMO-group weight in the two-bond shell. The last three refer to the sector fit. $D_{60}$ for one member of a doublet describes the particular real vector returned by the solver.}\label{tab:rotation}
\begin{tabular}{lrrrrrr}\toprule
Family & RMS (\AA) & Free fit (\%) & Sector fit (\%) & $g$ & $D_{60}$ single & $D_{60}$ group\\\midrule
I-ene & 0.01760 & 72.329 & 0.124 & 2 & 0.558 & $1.24\times10^{-8}$\\
I-yne & 0.02677 & 17.634 & 12.451 & 1 & $9.97\times10^{-10}$ & $9.97\times10^{-10}$\\
II-yne & 0.00397 & 4.757 & 4.753 & 2 & 0.554 & $1.23\times10^{-8}$\\
III-yne & $2.63\times10^{-6}$ & 9.005 & 9.005 & 2 & 0.553 & $1.19\times10^{-8}$\\\bottomrule
\end{tabular}\end{table}

The unconstrained I-ene geometry has an isolated, one-sided LUMO at the chosen resolution, with $D_{60}=0.996$ (geometry comparison in Table~\ref{tab:rotation}). Figure~\ref{fig:lumo_rotation} compares populations on the sector-fitted coordinates alone. For I-ene, an individual member of the LUMO doublet has $D_{60}=0.558$ and the complete doublet has $D_{60}=1.24\times10^{-8}$. Its small shell fraction, 0.124\%, is shared by both members. The edge-rich doublet in Table~\ref{tab:sector-frontier} lies higher in energy. Comparing the largest shell weights on the two structures tests boundary enrichment without assuming that the selected energy levels track continuously between geometries.

\begin{table}[htbp]\centering\small
\caption{Frontier levels of the sector-fitted G3 fragments. $\Delta$ is the HOMO--LUMO spacing. The shell column is the percentage of atoms in the two-bond mask. HOMO, LUMO and selected columns are mean group populations in that mask. $E_{\rm sel}$ is the selected group energy relative to the neutral midpoint. The selection excludes groups touching either end of the computed window.}\label{tab:sector-frontier}
\begin{tabular}{lrrrrrrr}\toprule
Family & C atoms & $\Delta$ (meV) & Shell (\%) & HOMO (\%) & LUMO (\%) & Selected (\%) & $E_{\rm sel}$ (meV)\\\midrule
I-ene & 13,950 & 2.346 & 7.96 & 0.157 & 0.124 & 79.05 & $+239.81$\\
I-yne & 21,774 & 0.069 & 6.14 & 18.62 & 12.45 & 19.41 & $+4.61$\\
II-yne & 19,590 & 3.408 & 6.22 & 9.04 & 4.75 & 10.42 & $+5.68$\\
III-yne & 22,140 & 5.761 & 5.80 & 12.45 & 9.00 & 25.66 & $-20.18$\\\bottomrule
\end{tabular}\end{table}

\section{Data and reproduction}\label{combined-si-sec-11}
\texttt{data/energies.csv} stores cell totals, and \texttt{code/update\_results.py} generates the tables. Periodic files in \texttt{data/bands/} and \texttt{data/dos/} use the full composition--linkage--alphabet name. The band JSON files give the actual input geometry, reciprocal path, DOS mesh and energy reference. The original electronic tables in \path{data/levels/} and \path{data/ldos/} supply G1, G2 and the unconstrained G3 comparison. The sector-fit calculation is in \path{sector_fit/}. Its \path{results/summary/} directory records the full 943-site sequence, bond-length residuals and group populations. Each family directory contains a Cartesian POSCAR and the six original atom indices associated with each independent position. The primary G3 levels and atomic weights are in \path{sector_fit/results/spectra/}. The separately distributed G3 calculation archive has the same layout from \path{results/} down. Geometry, transfer and spectral scripts accompany the data. The first and last energies for the four paired relaxations and the geometric differences are regenerated by \path{code/check_transfer.py}. \path{data/transfer_checks/} stores the three A7B8C12 inputs, outputs in coordinate form and energy traces. \path{data/mesh_spectra/} records the mesh and path eigenvalues used for the periodic spectra.
\section{Supplementary figures}\label{combined-si-sec-12}
Figures S1--S7 document the construction and selected atomic geometries. Figures S8--S13 give the computed spectra and boundary masks. Figure S14 compares seven predicted energies with relaxed DFT references. Figure S15 gives the metallic-mean comparison. Figures S16--S18 show the four G3 assemblies, their distance-transfer errors and the LUMO rotation test.
\clearpage
\newcommand{\sifigure}[3]{\begin{figure}[p]\centering\includegraphics[width=.96\linewidth,height=.79\textheight,keepaspectratio]{figs/#1.pdf}\caption{\textbf{#2} #3}\end{figure}}
\sifigure{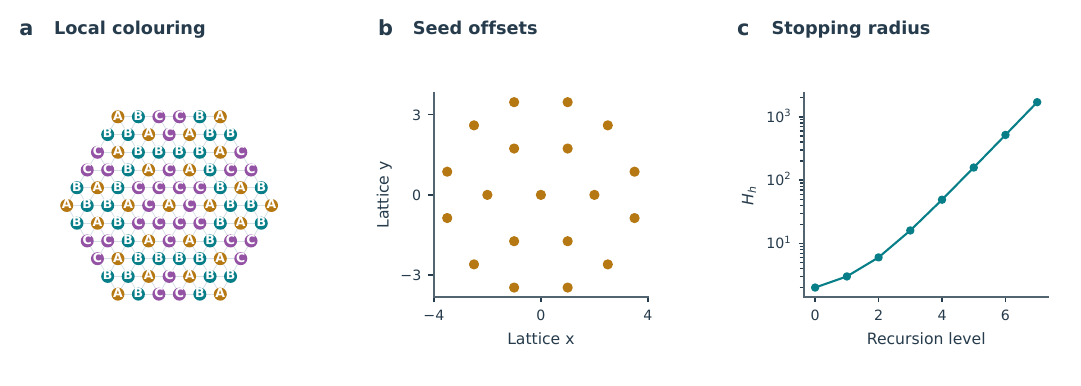}{Integer colouring and finite stopping.}{The local A/C/B field, the 19 digit offsets at the first scale, and the increasing lower bound $H_h$ on the distance of a nonzero high digit. These are exact construction diagrams; they are independent of chemical bond lengths.}
\sifigure{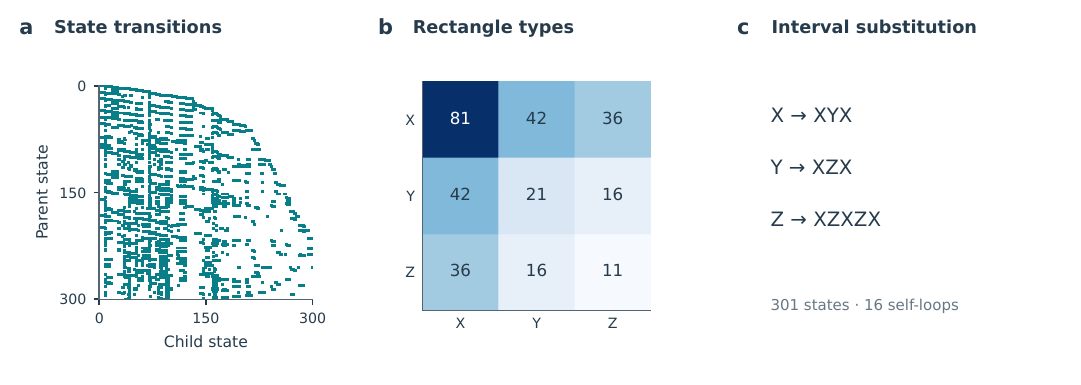}{Finite-state certificate.}{Sparsity of the 301-state transition matrix, the distribution of state rectangle types, and the interval substitution. The rectangle product is a partition device; the two-dimensional colour field is not assumed to factor into independent one-dimensional colours.}
\sifigure{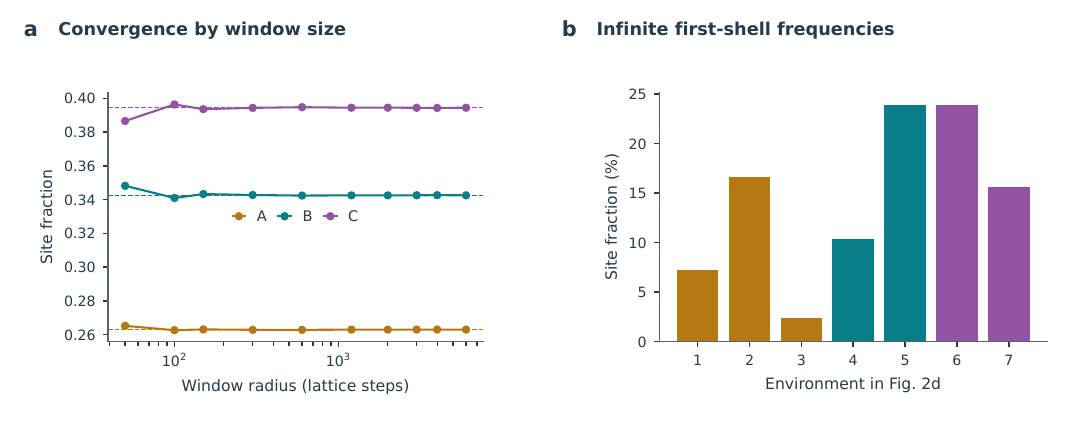}{Finite-window convergence and infinite frequencies.}{Centred hexagonal windows approach the exact molecular fractions. Horizontal dashed lines and the heights of the seven-environment bars denote exact limits. Numerical window counts corroborate the certificate and are not used to fit its frequencies.}
\sifigure{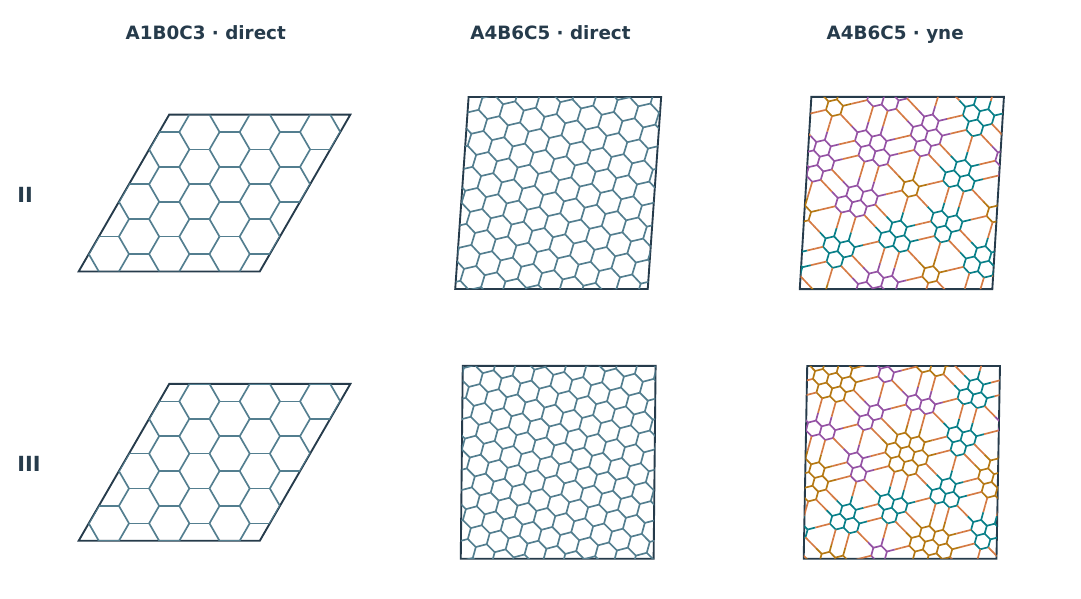}{Graphene recovered by direct benzenoid assembly.}{Representative II-ene and III-ene cells reduce to ordinary graphene. Inserting ethynylene connections retains the molecular partition in II-yne and III-yne. The direct assemblies are drawn in a single colour. The yne panels retain the construction-block colours.}
\sifigure{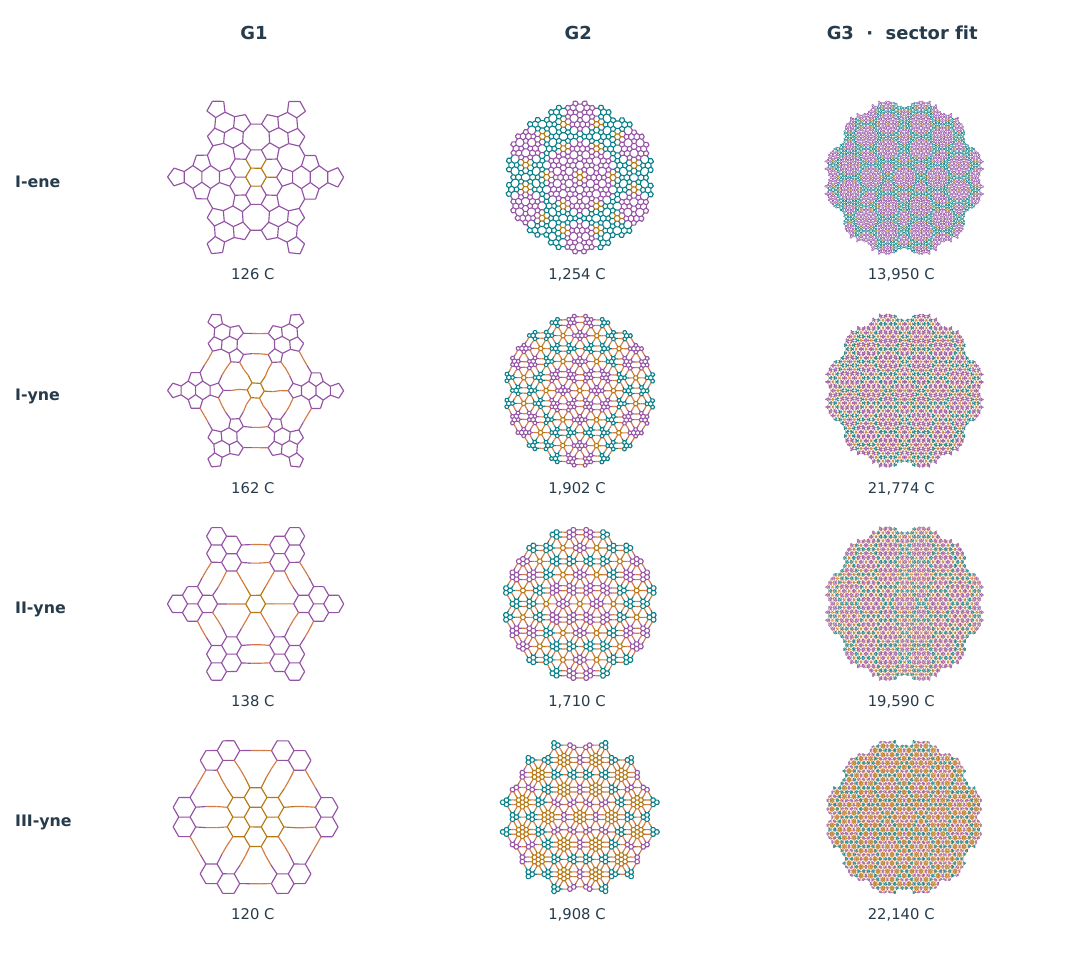}{Four chemical decorations on matched finite supports.}{G1 and G2 use the fitted carbon coordinates. Each G3 is assembled from one independent sector of the complete ABC code. Numbers count carbon atoms in the open fragments.}
\sifigure{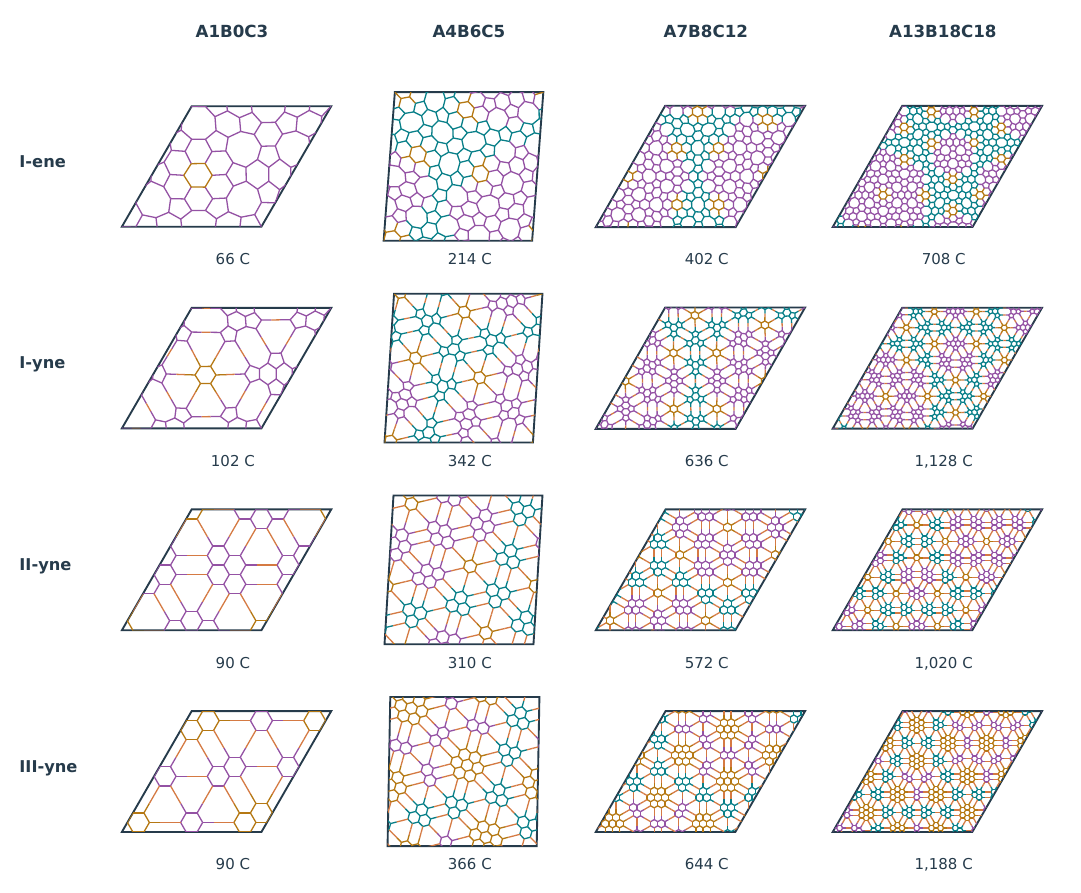}{Selected periodic atomic structures.}{The three basis cells and A13B18C18 holdout for each system. Cell boundaries and periodic bonds are shown using the archived atomic coordinates; counts are carbon atoms, not molecular sites.}
\sifigure{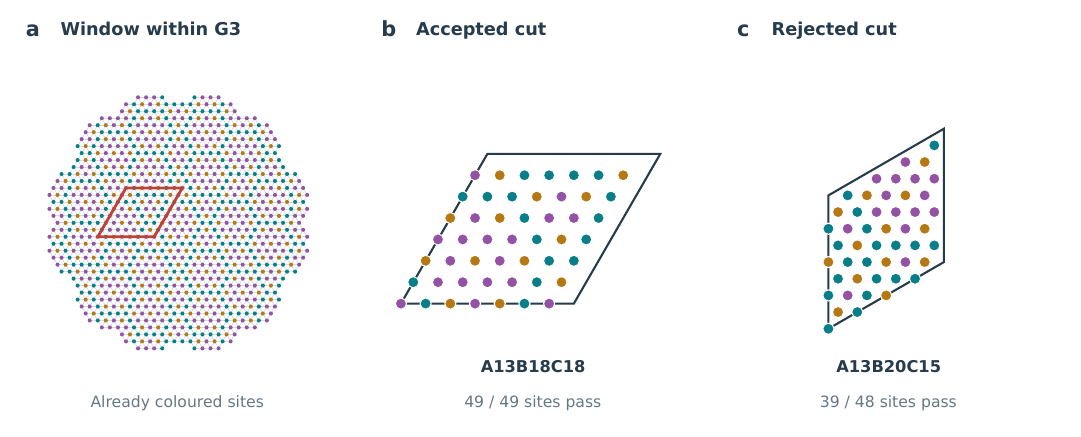}{A cut must pass its periodic seams.}{The accepted A13B18C18 fundamental window lies in the already coloured G3 support. Its periodic first shells all belong to the seven-type set. The A13B20C15 cut has nine disallowed centres and is excluded from this energy model.}
\sifigure{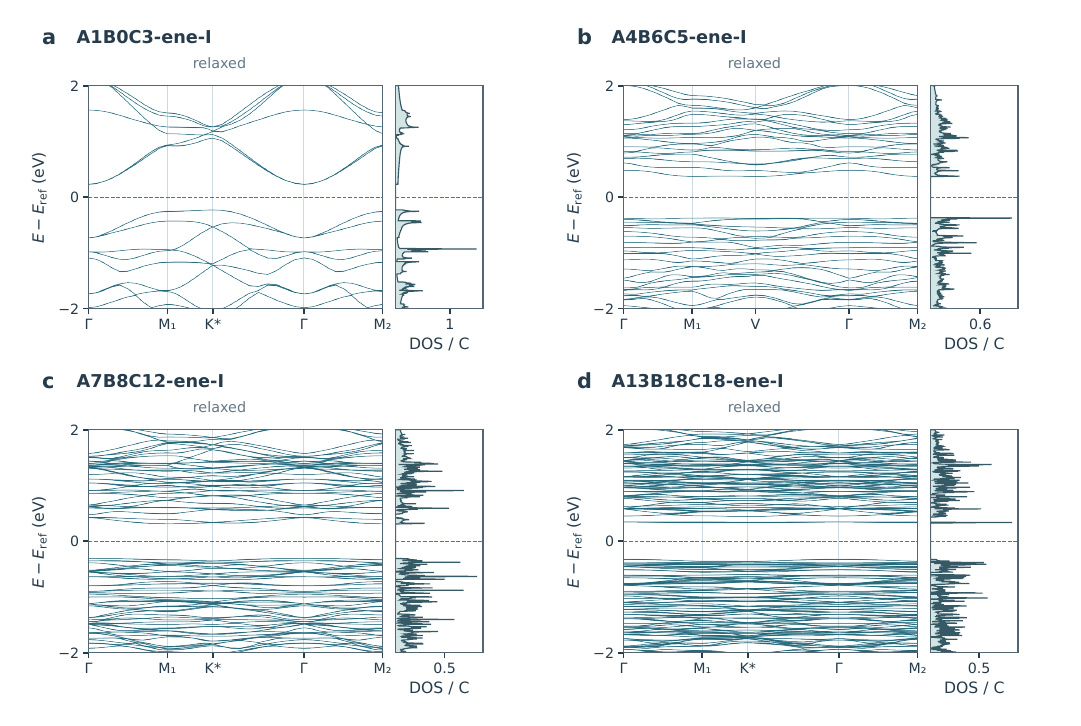}{I-ene periodic bands and DOS.}{All four underlying carbon geometries are relaxed.}
\sifigure{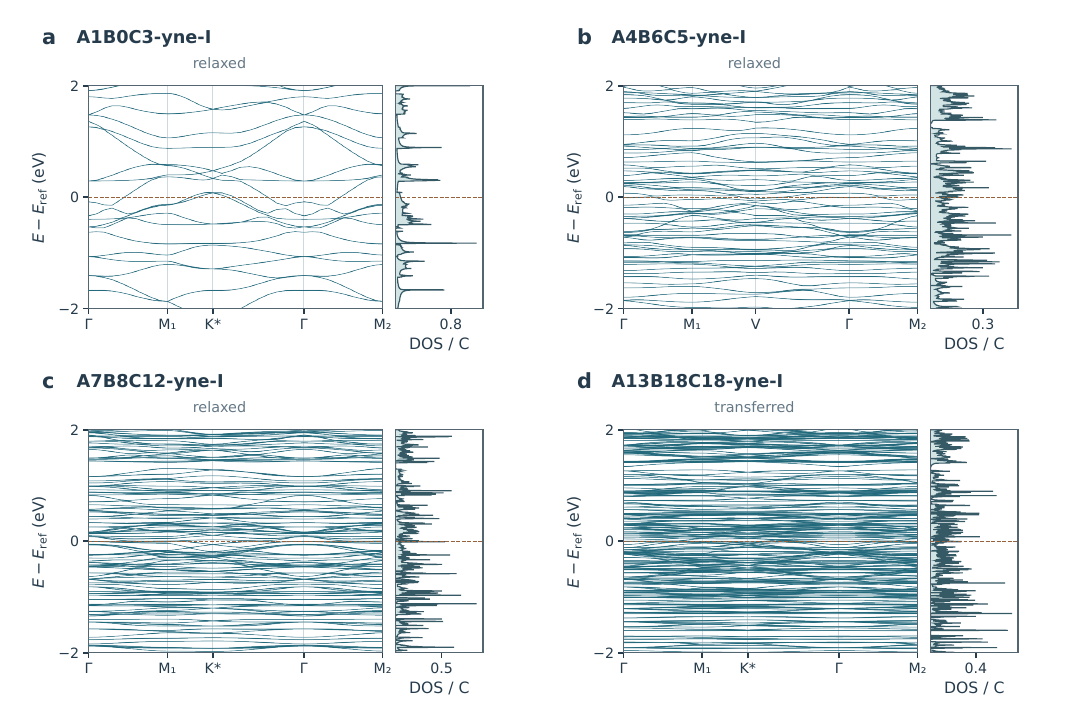}{I-yne periodic bands and DOS.}{The three basis geometries are relaxed. The large-cell curve uses transferred coordinates.}
\sifigure{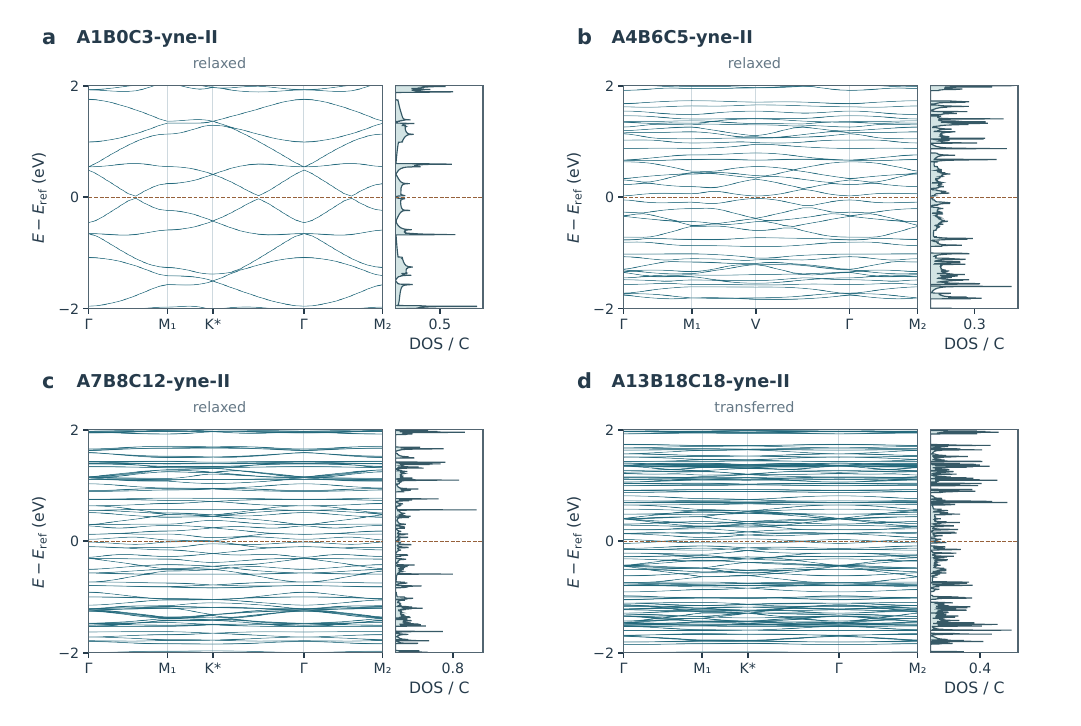}{II-yne periodic bands and DOS.}{The large cell uses transferred coordinates. DOS is spin summed per carbon.}
\sifigure{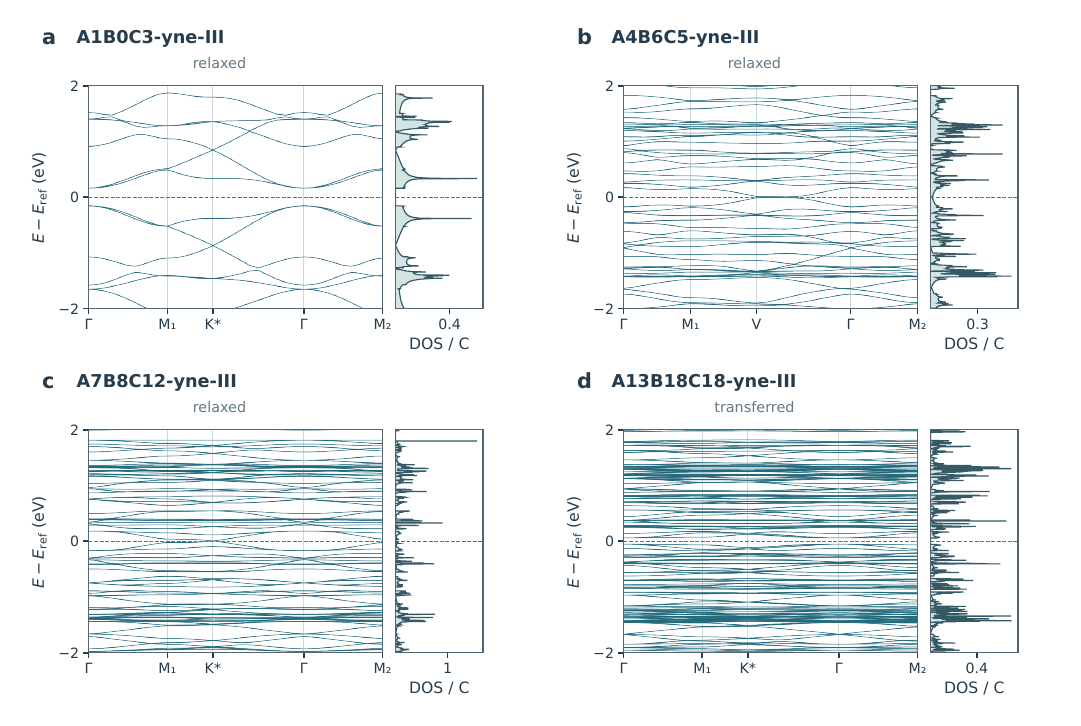}{III-yne periodic bands and DOS.}{The large cell uses transferred coordinates. In Figs.~S8--S11 each cell uses its own neutral reference and the same tight-binding model. All DOS curves are spin summed in states per eV per carbon.}
\sifigure{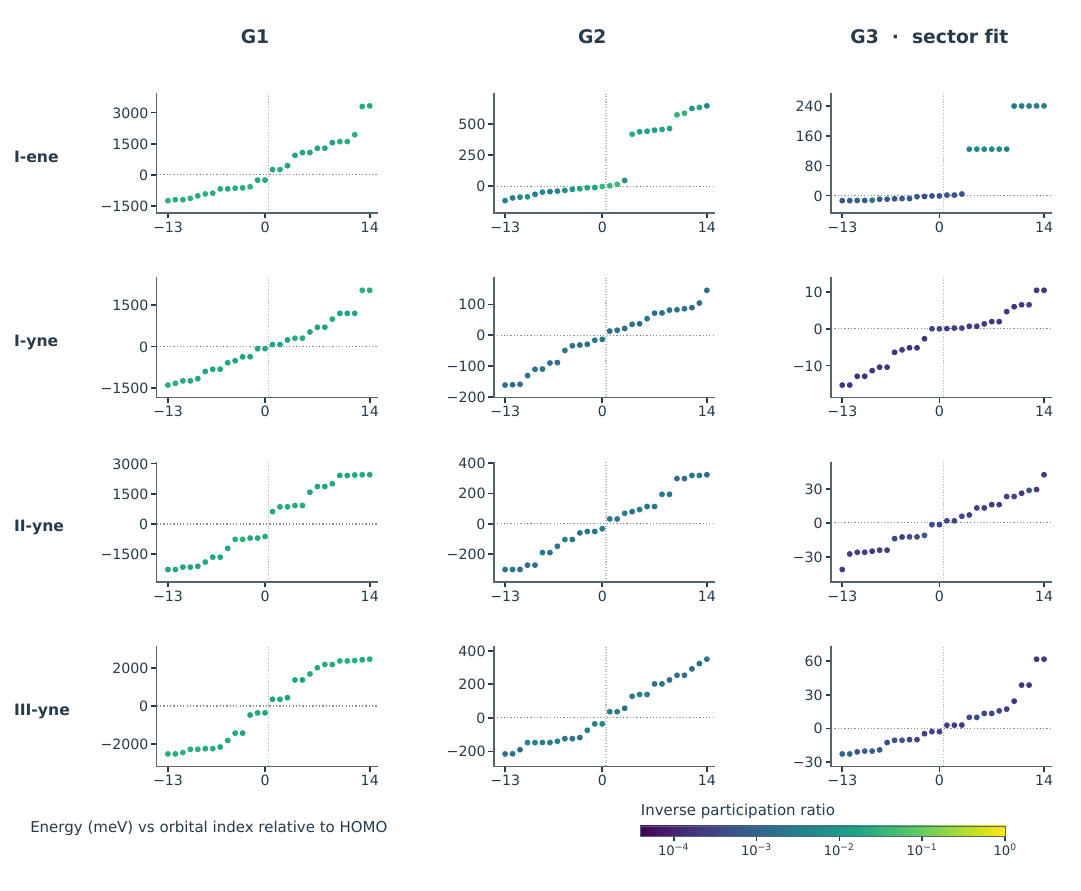}{Frontier levels across three generations.}{Each panel shows the 14 occupied and 14 unoccupied levels nearest the neutral midpoint in meV. Zero on the horizontal axis labels the HOMO and one labels the LUMO. Colours give the single-eigenvector inverse participation ratio. Each energy axis spans the displayed levels of that fragment. G3 uses the sector fit.}
\sifigure{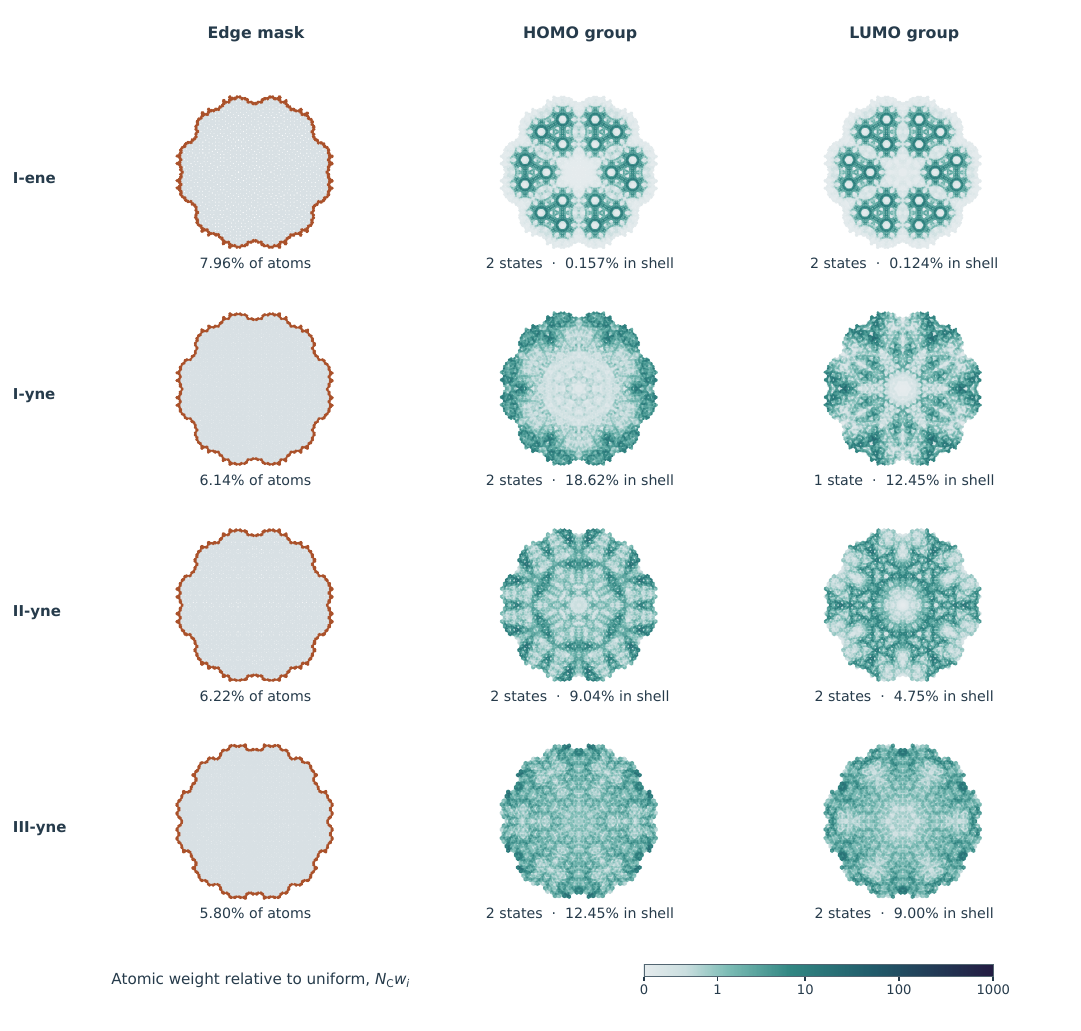}{Boundary masks and frontier populations of the sector-fitted G3 fragments.}{Left, the two-bond boundary shell. Middle and right, mean atomic populations of the HOMO and LUMO groups. Labels specify group size and shell weight. The same nonlinear $N_{\rm C}w_i$ scale applies to all populations. Horizontal carbon extents are 22.74, 31.04, 28.81 and 30.09 nm from top to bottom.}
\sifigure{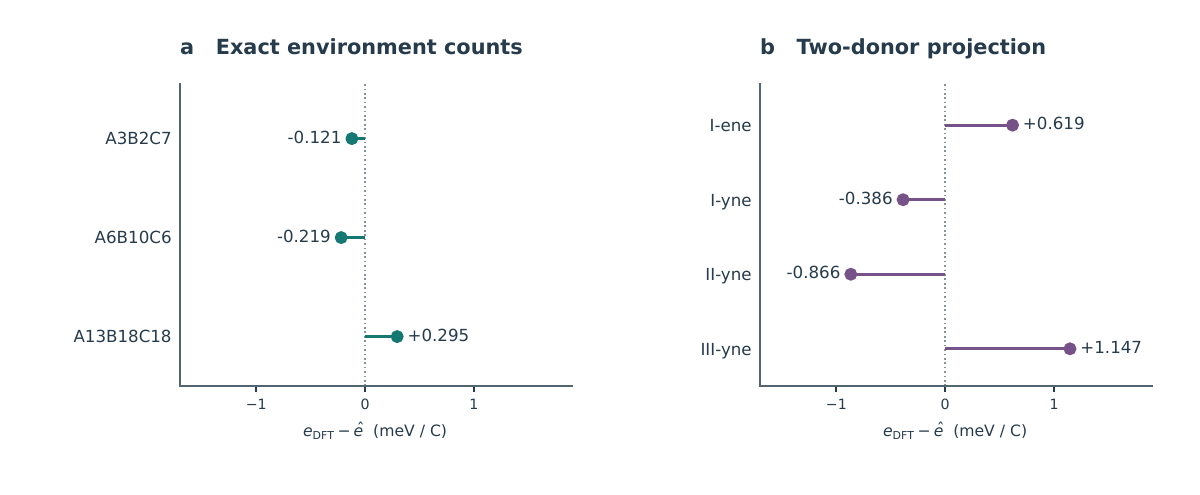}{Energy transfer across periodic cells.}{Signed DFT minus predicted energy per carbon. \textbf{a}, Three exact environment-count identities for I-ene. The small cells use the extended first-shell catalogue. \textbf{b}, The four chemical decorations of A7B8C12 predicted from the two smaller donors using the same projection of their site frequencies. Values and totals are listed in Table~\ref{tab:energytests}.}
\sifigure{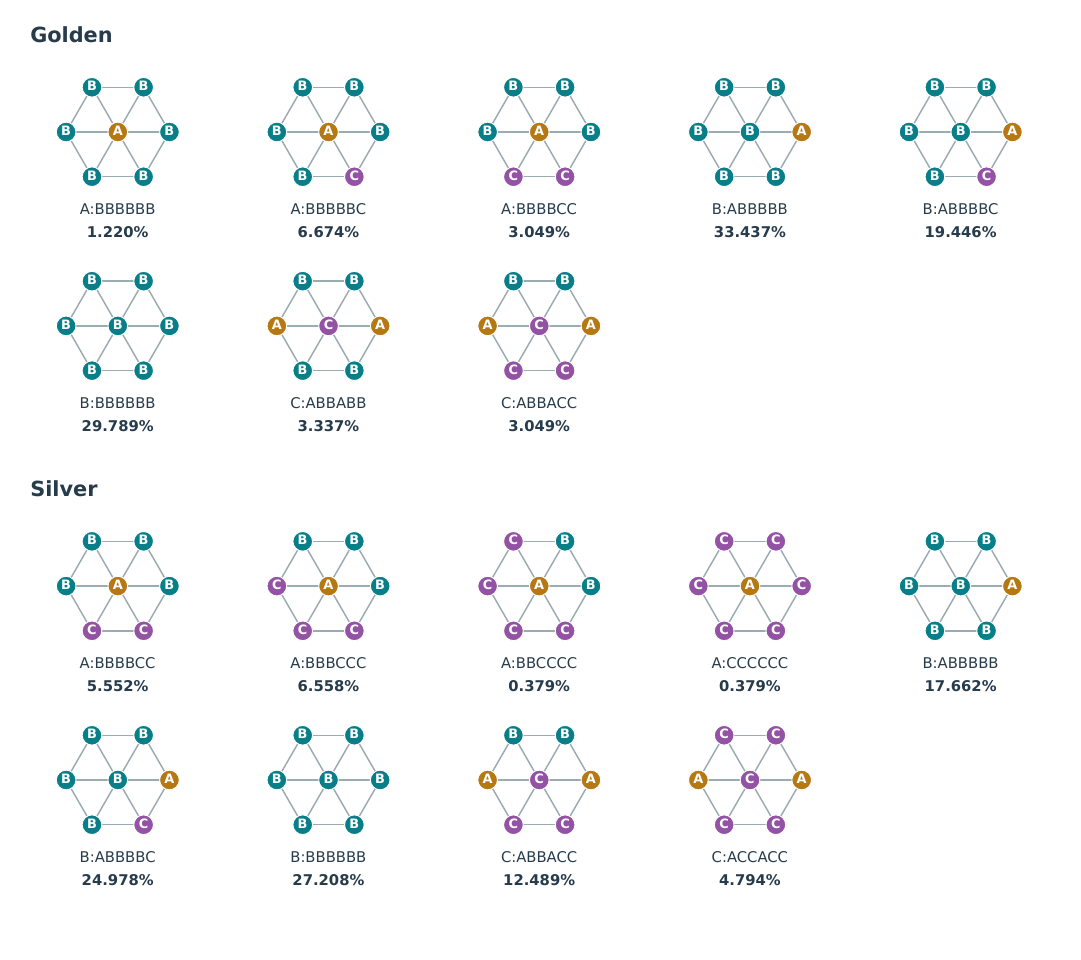}{Golden and silver nearest-neighbour catalogues.}{Each percentage is per lattice site in the infinite field. Rotations and reflections of the six neighbours are identified.}
\sifigure{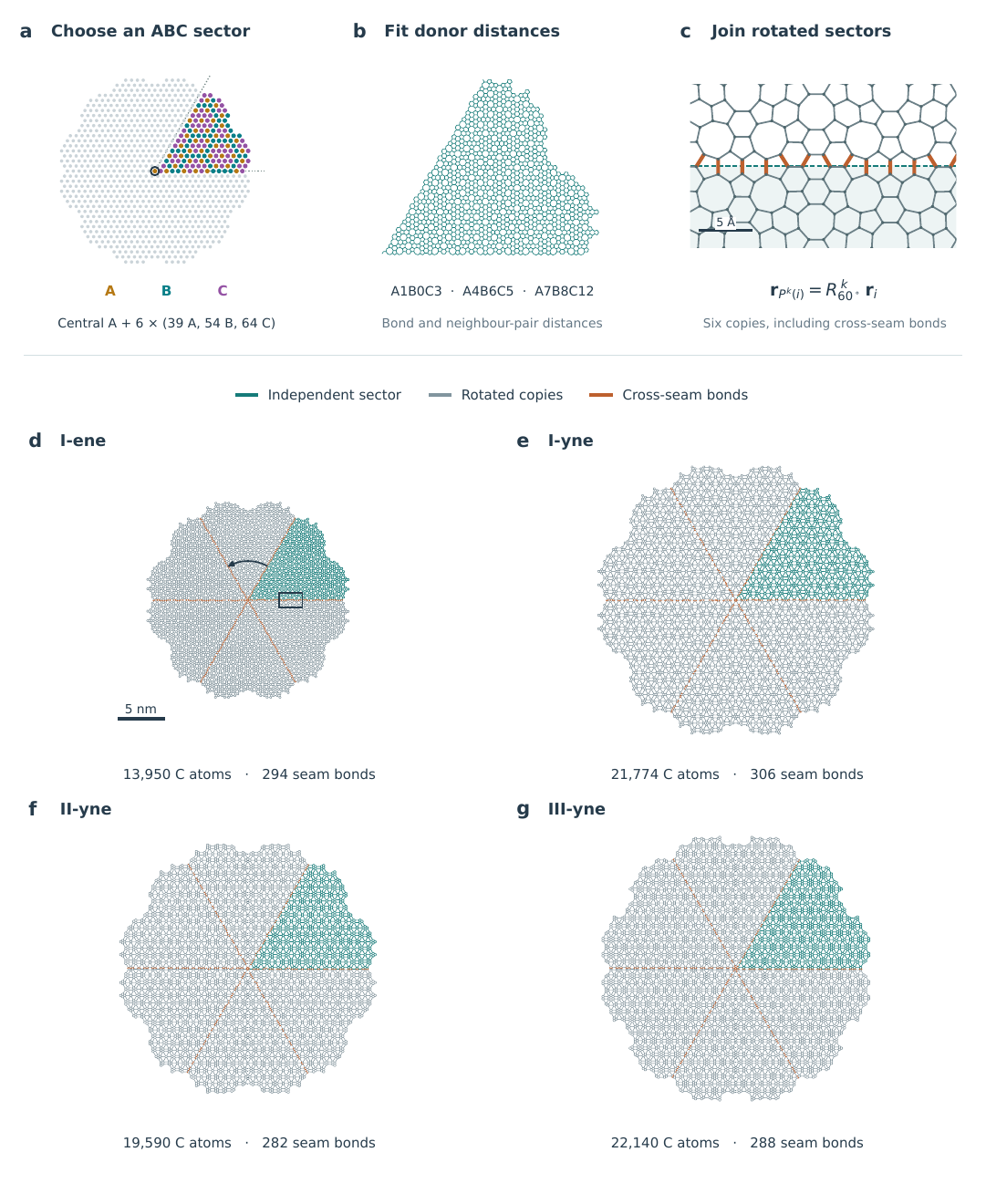}{Sixfold sector assembly on the complete G3 sequence.}{\label{fig:assembly}\textbf{a}, The circled central A site is shared. The coloured sector adds 39 A, 54 B and 64 C sites. \textbf{b}, Independent carbon positions for I-ene. Within each family, relaxed A1B0C3, A4B6C5 and A7B8C12 cells supply both bond lengths and distances between neighbours of a common carbon. \textbf{c}, The I-ene region boxed in \textbf{d}. The fit includes bonds joining adjacent sectors under the $60^\circ$ rotation, as in Eq.~\eqref{eq:sector-fit}. \textbf{d--g}, The complete I-ene, I-yne, II-yne and III-yne carbon bond graphs, drawn to the common 5~nm scale bar in \textbf{d}. The arrow marks a $60^\circ$ rotation. The highlighted sector contains all independent carbon coordinates. Each count of seam bonds covers all six boundaries.}
\sifigure{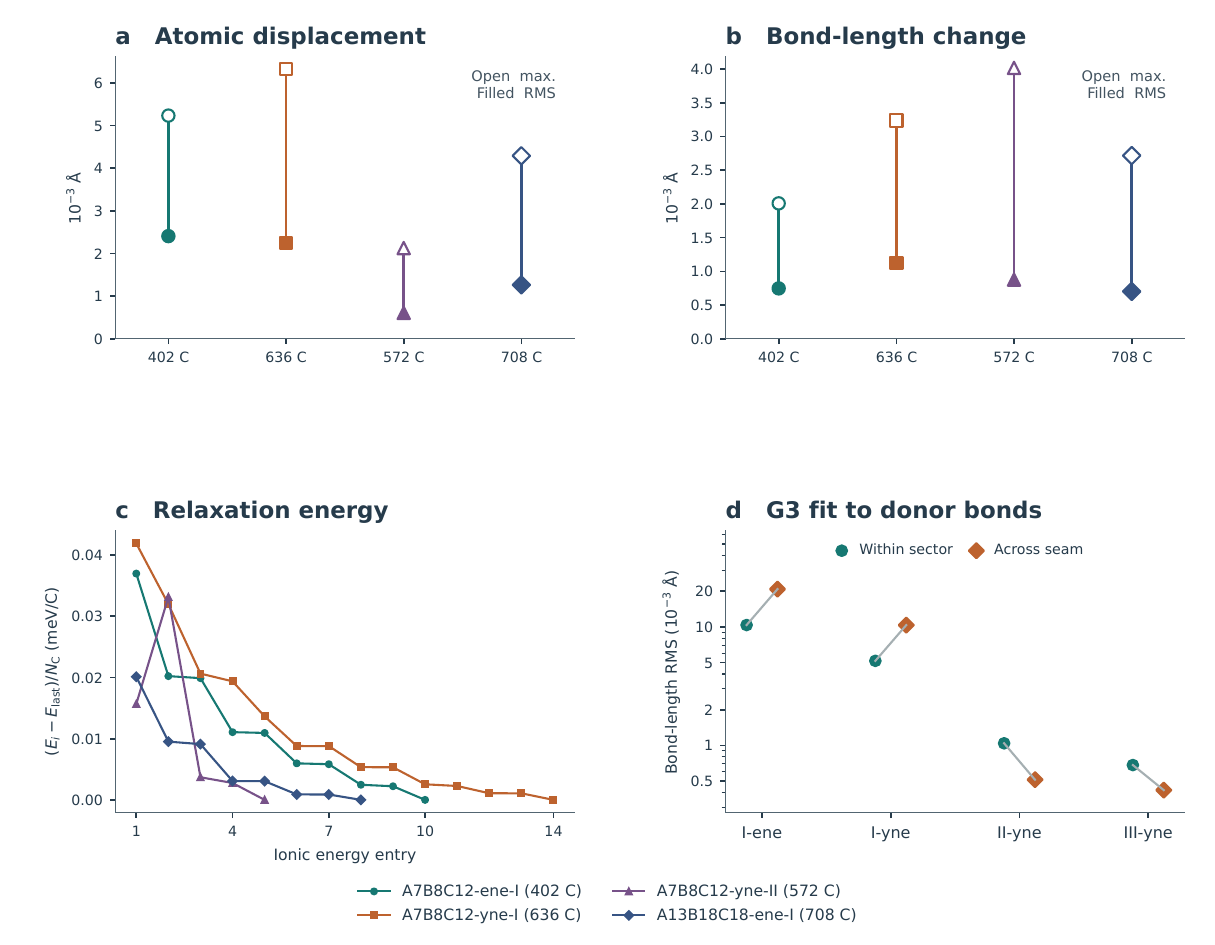}{Geometric residuals of assembled periodic cells and G3 fragments.}{\label{fig:residuals}\textbf{a}, Internal atomic shifts after periodic unwrapping, removal of homogeneous in-plane deformation and mean translation. \textbf{b}, Actual C--C bond-length changes including cell deformation. Solid and open symbols give RMS and maximum values, respectively. Numbers on the axes identify the four periodic cells by their carbon count. \textbf{c}, Total energy relative to the last entry of the same relaxation. \textbf{d}, G3 RMS bond-length deviations from the fitted donor targets for internal and cross-sector bonds. Table~\ref{tab:seams} gives their counts. The paired periodic geometries are quantified in Table~\ref{tab:paired}.}
\sifigure{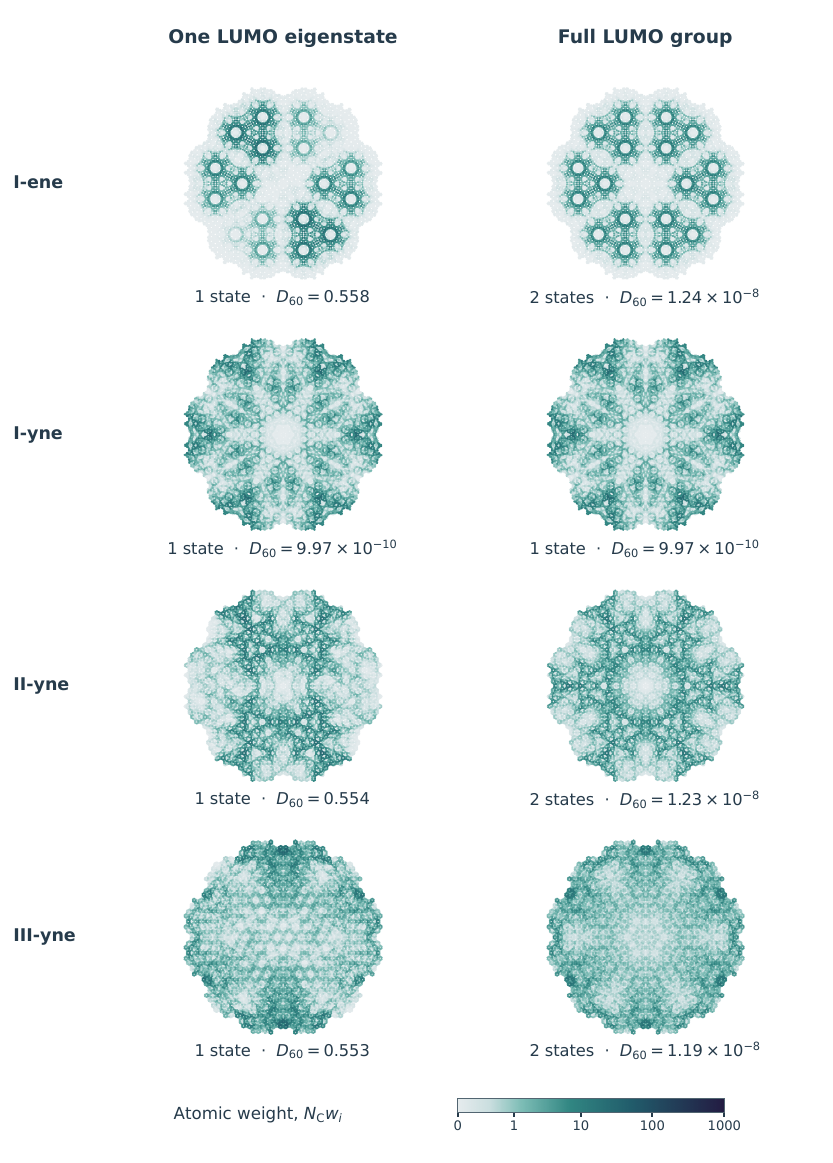}{Single-state and group LUMO populations on sixfold G3 fragments.}{\label{fig:lumo_rotation}Each row uses the same sector-fitted coordinates in both columns. Left, the atomic population of one LUMO eigenvector. Right, the mean population of the full LUMO group. The doublets in I-ene, II-yne and III-yne recover sixfold symmetry upon averaging. For the isolated I-yne LUMO, both columns depict the same state. The labels give the state count and the mismatch $D_{60}$ under a $60^\circ$ rotation. All panels use a common population scale. Table~\ref{tab:rotation} compares the unconstrained and sector-fitted geometries.}
\clearpage

\clearpage
\bibliographystyle{unsrtnat}
\bibliography{ref}

\begin{thebibliography}{26}
\providecommand{\natexlab}[1]{#1}
\providecommand{\url}[1]{\texttt{#1}}
\expandafter\ifx\csname urlstyle\endcsname\relax
  \providecommand{\doi}[1]{doi: #1}\else
  \providecommand{\doi}{doi: \begingroup \urlstyle{rm}\Url}\fi

\bibitem[Shechtman et~al.(1984)Shechtman, Blech, Gratias, and
  Cahn]{Shechtman1984}
D.~Shechtman, I.~Blech, D.~Gratias, and J.~W. Cahn.
\newblock Metallic phase with long-range orientational order and no
  translational symmetry.
\newblock \emph{Physical Review Letters}, 53:\penalty0 1951--1953, 1984.
\newblock \doi{10.1103/PhysRevLett.53.1951}.

\bibitem[Levine and Steinhardt(1984)]{Levine1984}
Dov Levine and Paul~Joseph Steinhardt.
\newblock Quasicrystals: A new class of ordered structures.
\newblock \emph{Physical Review Letters}, 53:\penalty0 2477--2480, 1984.
\newblock \doi{10.1103/PhysRevLett.53.2477}.

\bibitem[Levine and Steinhardt(1986)]{Levine1986}
Dov Levine and Paul~J. Steinhardt.
\newblock Quasicrystals. i. definition and structure.
\newblock \emph{Physical Review B}, 34:\penalty0 596--616, 1986.
\newblock \doi{10.1103/PhysRevB.34.596}.

\bibitem[Penrose(1974)]{Penrose1974}
Roger Penrose.
\newblock The role of aesthetics in pure and applied mathematical research.
\newblock \emph{Bulletin of the Institute of Mathematics and its Applications},
  10:\penalty0 266--271, 1974.

\bibitem[Dotera et~al.(2017)Dotera, Bekku, and Ziherl]{Dotera2017}
Tomonari Dotera, Shinichi Bekku, and Primoz Ziherl.
\newblock Bronze-mean hexagonal quasicrystal.
\newblock \emph{Nature Materials}, 16:\penalty0 987--992, 2017.
\newblock \doi{10.1038/nmat4963}.

\bibitem[Socolar and Taylor(2011)]{Socolar2011}
Joshua E.~S. Socolar and Joan~M. Taylor.
\newblock An aperiodic hexagonal tile.
\newblock \emph{Journal of Combinatorial Theory, Series A}, 118:\penalty0
  2207--2231, 2011.
\newblock \doi{10.1016/j.jcta.2011.05.001}.

\bibitem[Matsubara et~al.(2024)Matsubara, Koga, Takano, Matsushita, and
  Dotera]{Matsubara2024}
Toranosuke Matsubara, Akihisa Koga, Atsushi Takano, Yushu Matsushita, and
  Tomonari Dotera.
\newblock Aperiodic approximants bridging quasicrystals and modulated
  structures.
\newblock \emph{Nature Communications}, 15:\penalty0 5742, 2024.
\newblock \doi{10.1038/s41467-024-49843-4}.

\bibitem[Matsubara et~al.(2025)Matsubara, Koga, and Dotera]{Matsubara2025}
Toranosuke Matsubara, Akihisa Koga, and Tomonari Dotera.
\newblock Triangular and dice quasicrystals modulated by generic
  one-dimensional aperiodic sequences.
\newblock \emph{Physical Review B}, 111:\penalty0 104104, 2025.
\newblock \doi{10.1103/PhysRevB.111.104104}.

\bibitem[Fan et~al.(2021)Fan, Yan, Tripp, et~al.]{Fan2021}
Qitang Fan, Linghao Yan, Matthias~W. Tripp, et~al.
\newblock Biphenylene network: A nonbenzenoid carbon allotrope.
\newblock \emph{Science}, 372:\penalty0 852--856, 2021.
\newblock \doi{10.1126/science.abg4509}.

\bibitem[Toh et~al.(2020)Toh, Zhang, Lin, et~al.]{Toh2020}
Chee-Tat Toh, H.~Zhang, J.~Lin, et~al.
\newblock Synthesis and properties of free-standing monolayer amorphous carbon.
\newblock \emph{Nature}, 577:\penalty0 199--203, 2020.
\newblock \doi{10.1038/s41586-019-1871-2}.

\bibitem[Li et~al.(2010)Li, Li, Liu, Guo, Li, and Zhu]{Li2010}
Guoxing Li, Yuliang Li, Huibiao Liu, Yanbing Guo, Yongjun Li, and Daoben Zhu.
\newblock Architecture of graphdiyne nanoscale films.
\newblock \emph{Chemical Communications}, 46:\penalty0 3256--3258, 2010.
\newblock \doi{10.1039/B922733D}.

\bibitem[Ahn et~al.(2018)Ahn, Moon, Kim, et~al.]{Ahn2018}
Sung~Joon Ahn, Pilkyung Moon, Tae-Hoon Kim, et~al.
\newblock Dirac electrons in a dodecagonal graphene quasicrystal.
\newblock \emph{Science}, 361:\penalty0 782--786, 2018.
\newblock \doi{10.1126/science.aar8412}.

\bibitem[Yao et~al.(2018)Yao, Wang, Bao, et~al.]{Yao2018}
Wei Yao, Eryin Wang, Changhua Bao, et~al.
\newblock Quasicrystalline 30-degree twisted bilayer graphene as an
  incommensurate superlattice with strong interlayer coupling.
\newblock \emph{Proceedings of the National Academy of Sciences}, 115:\penalty0
  6928--6933, 2018.
\newblock \doi{10.1073/pnas.1720865115}.

\bibitem[Yu et~al.(2019)Yu, Wu, Zhan, Katsnelson, and Yuan]{Yu2019}
Guodong Yu, Zewen Wu, Zhen Zhan, Mikhail~I. Katsnelson, and Shengjun Yuan.
\newblock Dodecagonal bilayer graphene quasicrystal and its approximants.
\newblock \emph{npj Computational Materials}, 5:\penalty0 122, 2019.
\newblock \doi{10.1038/s41524-019-0258-0}.

\bibitem[Chernozatonskii et~al.(2022)Chernozatonskii, Demin, Kvashnin, and
  Kvashnin]{Diamane2022}
Leonid~A. Chernozatonskii, Victor~A. Demin, Alexander~G. Kvashnin, and
  Dmitry~G. Kvashnin.
\newblock Diamane quasicrystals.
\newblock \emph{Applied Surface Science}, 572:\penalty0 151362, 2022.
\newblock \doi{10.1016/j.apsusc.2021.151362}.

\bibitem[Wasio et~al.(2014)]{Molecular2014}
Natalie~A. Wasio et~al.
\newblock Self-assembly of hydrogen-bonded two-dimensional quasicrystals.
\newblock \emph{Nature}, 507:\penalty0 86--89, 2014.
\newblock \doi{10.1038/nature12993}.

\bibitem[Prodan and Kohn(2005)]{Prodan2005}
E.~Prodan and W.~Kohn.
\newblock Nearsightedness of electronic matter.
\newblock \emph{Proceedings of the National Academy of Sciences}, 102:\penalty0
  11635--11638, 2005.
\newblock \doi{10.1073/pnas.0505436102}.

\bibitem[van~de Walle and Ceder(2002)]{vandeWalle2002}
Axel van~de Walle and Gerbrand Ceder.
\newblock Automating first-principles phase diagram calculations.
\newblock \emph{Journal of Phase Equilibria}, 23:\penalty0 348--359, 2002.
\newblock \doi{10.1361/105497102770331596}.

\bibitem[Gong et~al.(2020)Gong, Shi, Li, Li, He, Ouyang, Zhang, Tang, and
  Zhong]{Gong2020}
Zhenhao Gong, Xizhi Shi, Jin Li, Shifang Li, Chaoyu He, Tao Ouyang, Chunxiao
  Zhang, Chao Tang, and Jianxin Zhong.
\newblock Theoretical prediction of low-energy {Stone--Wales} graphene with an
  intrinsic type-iii {Dirac} cone.
\newblock \emph{Physical Review B}, 101:\penalty0 155427, 2020.
\newblock \doi{10.1103/PhysRevB.101.155427}.

\bibitem[Shi et~al.(2021)Shi, Li, Li, Ouyang, Zhang, Tang, He, and
  Zhong]{He2021}
Xizhi Shi, Shifang Li, Jin Li, Tao Ouyang, Chunxiao Zhang, Chao Tang, Chaoyu
  He, and Jianxin Zhong.
\newblock High-throughput screening of two-dimensional planar sp2 carbon space
  associated with a labeled quotient graph.
\newblock \emph{The Journal of Physical Chemistry Letters}, 12:\penalty0
  11511--11519, 2021.
\newblock \doi{10.1021/acs.jpclett.1c03193}.

\bibitem[Shi et~al.(2018)Shi, He, Pickard, Tang, and Zhong]{Shi2018RG2}
Xizhi Shi, Chaoyu He, Chris~J. Pickard, Chao Tang, and Jianxin Zhong.
\newblock Stochastic generation of complex crystal structures combining group
  and graph theory with application to carbon.
\newblock \emph{Physical Review B}, 97:\penalty0 014104, 2018.
\newblock \doi{10.1103/PhysRevB.97.014104}.

\bibitem[Kresse and Furthmuller(1996)]{Kresse1996}
Georg Kresse and Jurgen Furthmuller.
\newblock Efficient iterative schemes for ab initio total-energy calculations
  using a plane-wave basis set.
\newblock \emph{Physical Review B}, 54:\penalty0 11169--11186, 1996.
\newblock \doi{10.1103/PhysRevB.54.11169}.

\bibitem[Solomyak(1997)]{Solomyak1997}
Boris Solomyak.
\newblock Dynamics of self-similar tilings.
\newblock \emph{Ergodic Theory and Dynamical Systems}, 17:\penalty0 695--738,
  1997.
\newblock \doi{10.1017/S0143385797084988}.

\bibitem[Solomyak(1999)]{Solomyak1999}
Boris Solomyak.
\newblock Corrections to `dynamics of self-similar tilings'.
\newblock \emph{Ergodic Theory and Dynamical Systems}, 19:\penalty0 1685--1685,
  1999.
\newblock \doi{10.1017/S014338579917161X}.

\bibitem[Blochl(1994)]{Blochl1994}
Peter~E. Blochl.
\newblock Projector augmented-wave method.
\newblock \emph{Physical Review B}, 50:\penalty0 17953--17979, 1994.
\newblock \doi{10.1103/PhysRevB.50.17953}.

\bibitem[Perdew et~al.(1996)Perdew, Burke, and Ernzerhof]{PBE1996}
John~P. Perdew, Kieron Burke, and Matthias Ernzerhof.
\newblock Generalized gradient approximation made simple.
\newblock \emph{Physical Review Letters}, 77:\penalty0 3865--3868, 1996.
\newblock \doi{10.1103/PhysRevLett.77.3865}.

\end{thebibliography}
\end{document}